\documentclass[preprintnumbers,showpacs,amsmath,amssymb,floatfix,9pt,prd,onecolumn,
superscriptaddress,nofootinbib]{revtex4}
\usepackage{graphicx}
\usepackage{latexsym}
\usepackage{epsfig}
\usepackage{amssymb}
\usepackage[utf8]{inputenc}
\begin{document}
\title{Isotropic Buchdahl's relativistic fluid sphere within $f(R,\,T)$ gravity}

\author{Piyali Bhar \footnote{Corresponding author}}
\email{piyalibhar90@gmail.com , piyalibhar@associates.iucaa.in}
\affiliation{Department of
Mathematics, Government General Degree College, Singur, Hooghly, West Bengal 712409,
India}

\author{Pramit Rej }
\email{pramitrej@gmail.com
 } \affiliation{Department of
Mathematics, Sarat Centenary College, Dhaniakhali, Hooghly, West Bengal 712 302, India}

\begin{abstract}
The aim of the research is to look into a new solution for isotropic compact stars in the context of the $f(R,\,T)$ theory of gravity. We used the Buchdahl [H.A. Buchdahl, Phys. Rev. {\bf 116} (1959) 1027] metric potentials as input to deal with the field equations in the $f(R,\,T)$ framework. For different values of the coupling parameter $\chi$, graphical representation of the model parameters have been shown to canvass the analytical results more clearly. Interestingly, we have proven that for $\chi=0$, the standard General Relativity (GR) results can be recovered. A comparison of our obtained solutions with the GR results is also discussed. To study the effect of the coupling parameter $\chi$, the numerical values of the different physical variables have been tabulated for the values of the coupling parameter $\chi=0,\,0.25,\,0.5,\,0.75,1,\,1.25$. We used the compact stars candidate LMC X-4 with mass$=(1.04 \pm 0.09)M_{\odot}$; Radius $= 8.301_{-0.2}^{+0.2}$ km. respectively, for graphical analysis. To determine the physical acceptability of the model, we looked into the necessary physical properties such as energy conditions, causality, hydrostatic equilibrium, and pressure-density ratio etc. and found that our system satisfies all of these criteria, indicating that the model is physically reasonable.
\end{abstract}



\maketitle

\section{Introduction}

Massive stars explode as supernova to the end of their life and yield extremely compact objects with an average density of $10^{14}$ gm.cm$^{-3}$. The internal matter of these compact objects is compressed by the strong gravitational fields to densities that range from sub-saturation to a few times nuclear saturation density, $n_0 =$ 0.16 fm$^{-3}$ \cite{glendenning2012compact}. In $1934$, Baade and Zwicky \cite{baade1934remarks} set up the idea that massive compact stellar objects could form, establishing the theory that a supernova might produce a small, super dense star.

There are two distinct theories that might be used to settle the argument over how to account for the universe's accelerating expansion. One is the possibility that mysterious dark energy exists, as well as its possible expansions, such as modified gravity theories. The cosmological constant, which represents a constant energy density of the vacuum and satisfies cosmological data, is the simplest illustration of dark energy. This problematic nature of
cosmological constant has motivated intense research for alternative theories of gravity extending the Einstein's
theory of gravity. This leads to the search for a different gravity theory that can answer the universe's current acceleration phase. Alternative explanations have been demonstrated to be capable of adequately describing cosmological observations. One of the simplest possible modification is the $f(R)$-gravity. Another alternative theory of gravity is so called $f(R,T)$ gravity. An interesting aspect of $f(R,T)$ theory is that it may provide an effective classical description of the quantum properties of gravity. In addition to improving fundamental understanding, this theory has produced certain results. The other motivations are related to reconstructing $f(R,T)$ gravity from holographic dark energy, cosmological and solar System Consequences, anisotropic cosmology, non-equilibrium picture of thermodynamics, a wormhole solution, and some other relevant aspects. However, it is vital to do astrophysical research, such as using relativistic stars, in order to develop a suitable gravity theory. Some justifications for these theories are based on the idea that relativistic stars in a strong gravitational field may distinguish between the fundamental laws of gravity and its generalizations. Considering all the facts, we consider here $f(R,T)$ gravity theory from the set of alternative theories of gravity.
By considering modified theories of gravity like $f(R, T)$  gravity, the problem of accelerated expansion of the universe can be resolved \cite{Harko:2011kv}. The $f(R,\,T)$ gravity offers an alternate explanation for the current cosmic acceleration without requiring the introduction of either an exotic dark energy component or the creation of additional spatial dimensions. Cosmic acceleration in $f(R,\,T)$ gravity may be caused by matter contents in addition to geometrical contributions to the total cosmic energy density \cite{Zubair:2015gsb}.

The $f(R,T)$ theory has become increasingly popular among researchers in recent decades. Harko et al. \cite{Harko:2011kv} studied $f(R,T)$ modified theories of gravity, in which the gravitational Lagrangian is given by an arbitrary function of the Ricci scalar $R$ and of the trace of the stress-energy tensor $T$, and obtained gravitational field equations in the metric formalism, as well as equations of motion for test particles, which follow from the covariant divergence of the stress-energy tensor. By choosing $f(R,\,T)$ as a linear function of $R$ and $T$, Errehymy et al. \cite{Errehymy:2022ihw} investigated the existence of compact structures describing anisotropic matter distributions within the framework of $f(R,\,T)$ theory. They used the embedding class one technique to obtain a full space-time description inside the stellar configuration. Sarkar et al. \cite{Sarkar:2022vko} proposed a completely new model for spherically symmetric anisotropic compact stars with class one solutions in the context of $f(R,\,T)$ gravity, with all physical parameters derived by using a suitable $g_{rr}$ metric function.
In the paradigm of $f(R,\,T)$ gravity theories, Tangphati et al. \cite{Tangphati:2022mur} examined compact static configurations whose matter field is composed of homogenous, neutral 3-flavor interacting quark matter with O(ms4) corrections, namely, interacting quark EoS. Azmat et al. \cite{Azmat:2022bbc} used gravitational decoupling and the minimal geometric deformation (MGD) approach to build an analytical version of the gravastar model with non-uniform and anisotropic features in the framework of $f(R,\,T)$ gravity. Bhattacharjee \cite{Bhattacharjee:2022lcs} uses mimetic $f(R,\,T)$ gravity, a Lagrange multiplier, and a mimetic potential to produce feasible inflationary cosmological solutions that are consistent with the most recent Planck and BICEP2/Keck Array data. In the context of the $f(R,\,T)$ theory of gravity, Baffou et al. \cite{Baffou:2021ycm} examined the cosmological inflation scenario, Prasad et al. \cite{Prasad:2021yrf} study the existence of a compact star configuration model. They used a well-known Karmarkar condition to create a well-behaved embedding class-one solution with a specified linear function for $f(R,\,T)$, Yousaf \cite{Yousaf:2021tol} investigated axial and reflection-symmetric self-gravitating sources. Ahmed et al.\cite{Ahmed:2021fav} used a linear model for $f(R,\,T)$ gravity to examine the dissipative gravitational collapse of an anisotropic spherically symmetric radiating star that meets the initially static Karmarkar condition. Kumar et al. \cite{Kumar:2021vqa} proposed an isotropic compact star model in $f(R,\,T)$ gravity using Buchdahl anastz. By considering the conjecture of Mazur and Mottola in general relativity, Sharif and Waseem \cite{Sharif:2019oni} investigated the effects of charge on stellar objects, known as gravastars, under the influence of $f(R,\,T)$ gravity. Noureen et al. \cite{Noureen:2021xlf} investigated the evolution of spherically symmetric charged anisotropic viscous fluids in the context of $f(R,\,T)$ gravity and employed the perturbation scheme to analyse stability. Maurya et al. \cite{Maurya:2021aio} presented a new method for constructing self-gravitating systems rely on imperfect fluid distributions that is both simple and effective. This method was created within the context of the $f(R,\,T)$ gravity theory by integrating two geometrical schemes: gravitational decoupling via minimal geometric deformation and the embedding technique, specifically the class I grip.
Hansraj and Banerjee \cite{Hansraj:2018jzb} studied at the behavior of well-known stellar models in the frame of the $f(R,\,T)$ modified theory of gravity, and discovered that in some circumstances, the $f(R,\,T)$ model displays more pleasing behavior than its Einstein counterpart. Barrientos et al. \cite{Barrientos:2018cnx} investigated $f(R,\,T)$ gravity theories and discovered that once an effective energy-momentum tensor is applied, the resulting field equations are quite similar to their metric-affine $f(R)$ siblings. By choosing a minimal coupling between matter and gravity, Gamonal \cite{Gamonal:2020itt} investigated the slow-roll approximation to cosmic inflation in the setting of $f(R,\,T)$ gravity. In the $f(R,\,T)$ theory of gravity, Moraes \cite{Moraes:2015kka} gave accurate cosmological solutions derived from Wesson's induced matter model and applied to a general 5D metric.\\

The purpose of this study is to investigate the appearance of $f(R,\,T)$ gravity in modeling realistic configurations of compact stellar objects in the presence of the Buchdahal metric {\em ansatz}, as well as the discussion on the stability and physical characteristics of compact star LMC X-4. There are so many earlier works on stellar models within the context of f(R,T) gravity, and established that in some situation these theories displays more pleasing behavior than its Einstein GTR. Motivated by these good antecedents, here we extend Buchdahl's spacetime from GR framework to the f(R, T) gravity field. The Buchdahl ansatz is a well-known solutions in GR with a clear geometric characterization of the associated spacetime metric. In particular, this ansatz provides a model which is singularity free. In this paper, we looked into the necessary physical properties such as energy conditions, causality, hydrostatic equilibrium, and pressure-density ratio etc. and found that our system satisfies all of these criteria, indicating that the model is physically viable. Also we compare our results with the GR results. Our solutions provide a toy model that closely resembles the behavior of real astrophysical objects like neutron stars, white dwarfs, or strange star families. As a result, the knowledge gained from these celestial bodies has improved our understanding of, among other things, the behavior of gravitational interaction in the strong field regime and some complex processes of particle creation and annihilation. In the context of an isotropic matter source, we study various structural features by selecting a specific form of $f(R,\,T)$ gravity models. The Tolman-Oppenheimer-Volkoff (TOV) equation, mass radius relation, compactness parameter, surface redshift, stability, and various energy conditions will all be thoroughly investigated.\\
The following is the layout of this paper: The mathematical definition of $f(R,T)$ gravity in the context of isotropic matter distributions is presented in Section \ref{sec2}. Section \ref{sec3} shows some of the possible $f(R,\,T)$ gravity models and Section \ref{sec4} represents boundary conditions, here we match the interior metric to Schwarzschild's exterior metric to calculate the values of the unknown constant for the chosen values of our model parameters. Section \ref{sec5} examines some physical characteristics as well as the viability of some familiar compact stars using graphical analysis. The final section is focused on the concluding remarks.

\section{Interior Spacetime and Basic field Equations}\label{sec2}
Harko {\em et al.} \cite{Harko:2011kv} proposed the Einstein Hilbert action for $f(R,T )$ gravity which is given by,
\begin{eqnarray}\label{action}
S&=&\frac{1}{16 \pi}\int  f(R,T)\sqrt{-g} d^4 x + \int \mathcal{L}_m\sqrt{-g} d^4 x,
\end{eqnarray}
$f ( R,T )$ denotes the general function of  trace $T$ along with Ricci scalar $R$ and $\mathcal{L}_m$ being the lagrangian matter density and $g = det(g_{\mu \nu}$). According to Landau and Lifshitz \cite{landau2013classical}, the stress-energy tensor of matter is defined as,
\begin{eqnarray}\label{tmu1}
T_{\mu \nu}&=&-\frac{2}{\sqrt{-g}}\frac{\delta \sqrt{-g}\mathcal{L}_m}{\delta \sqrt{g_{\mu \nu}}},
\end{eqnarray}
and its trace is given by $T=g^{\mu \nu}T_{\mu \nu}$. If the Lagrangian density $\mathcal{L}_m$ depends only on $g_{\mu \nu}$, on on its derivatives, eqn.(\ref{tmu1}) becomes,
\begin{eqnarray}
T_{\mu \nu}&=& g_{\mu \nu}\mathcal{L}_m-2\frac{\partial \mathcal{L}_m}{\partial g_{\mu \nu}}.
\end{eqnarray}
The field equations of the $f(R,T)$ gravity corresponding to action (\ref{action}) is given by,
\begin{eqnarray}\label{frt}
f_R(R,T)R_{\mu \nu}-\frac{1}{2}f(R,T)g_{\mu \nu}+(g_{\mu \nu }\Box-\nabla_{\mu}\nabla_{\nu})f_R(R,T)&=&8\pi T_{\mu \nu}-f_T(R,T)T_{\mu \nu}\nonumber\\&&-f_T(R,T)\Theta_{\mu \nu}.
\end{eqnarray}
Where, $f_R(R,T)=\frac{\partial f(R,T)}{\partial R},~f_T(R,T)=\frac{\partial f(R,T)}{\partial T}$. $\nabla_{\nu}$ represents the covariant derivative
associated with the Levi-Civita connection of $g_{\mu \nu}$, $\Theta_{\mu \nu}=g^{\alpha \beta}\frac{\delta T_{\alpha \beta}}{\delta g^{\mu \nu}}$ and
$\Box \equiv \frac{1}{\sqrt{-g}}\partial_{\mu}(\sqrt{-g}g^{\mu \nu}\partial_{\nu})$ represents the D'Alambert operator.\\

Now the divergence of the stress-energy tensor $T_{\mu \nu}$ can be obtained by the taking covariant divergence of (\ref{frt}) (For details see ref \cite{Harko:2011kv} and\cite{koivisto2006note}) as,
\begin{eqnarray}\label{conservation}
\nabla^{\mu}T_{\mu \nu}&=&\frac{f_T(R,T)}{8\pi-f_T(R,T)}\left[(T_{\mu \nu}+\Theta_{\mu \nu})\nabla^{\mu}\ln f_T(R,T)+\nabla^{\mu}\Theta_{\mu \nu}\right].
\end{eqnarray}

From eqn.(\ref{conservation}), we can check that $\nabla^{\mu}T_{\mu \nu}\neq 0$ if $f_T(R,T)\neq 0.$ So like Einstein gravity, the system will not be conserved. It can be noted that when $f(R,T)=f(R)$, from eqn. (\ref{frt}) we obtain
the field equations of $f(R)$ gravity.\par

In curvature coordinates $(t,r,\theta,\phi)$,
\begin{equation}\label{line}
ds^{2}=-e^{\nu}dt^{2}+e^{\lambda}dr^{2}+r^{2}d\Omega^{2},
\end{equation}
provides the static and spherically symmetric line element, where $d\Omega^{2}\equiv \sin^{2}\theta d\phi^{2}+d\theta^{^2}$ and the metric co-efficients $\nu$ and $\lambda$ purely radial functions.
In the current work, we make the assumption that the fluid around a compact star is perfect. Consequently, the stress-energy tensor of matter is given by,
\begin{eqnarray}
T_{\mu \nu}&=&(p+\rho)u_{\mu}u_{\nu}-p g_{\mu \nu},
\end{eqnarray}
where $\rho$ is the matter density, $p$ is the isotropic pressure in modified gravity, $u^{\mu}$ is the fluid four velocity satisfies the equations $u^{\mu}u_{\mu}=1$ and $u^{\mu}\nabla_{\nu}u_{\mu}=0$ and following Harko et al.\cite{Harko:2011kv} the matter Lagrangian can be taken as $\mathcal{L}_m=-p$ and the expression of $\Theta_{\mu \nu}=-2T_{\mu \nu}-pg_{\mu\nu}.$\\

Let's take a separable functional form given by,
\begin{eqnarray}
f (R, T ) = f_1(R)+f_2(T ),
\end{eqnarray}
in the context of relativistic structures to discuss the coupling effects of matter and curvature components in $f(R,\, T)$ gravity.
Where $f_1(R)$ and $f_2(T)$ being arbitrary functions of $R$ and $T$ respectively. By selecting several $f_1(R)$ forms and combining them linearly with $f_2(T)$ in $f(R,\,T)$ gravity, several feasible models can be produced. We take into account $f_1(R)=R$ and $f_2(T)=2\chi T $ in our current model. i.e., we choose
\begin{eqnarray}\label{e}
f(R,T)&=& R+2 \chi T,
\end{eqnarray}
$\chi$ is some constant.
Using (\ref{e}) into (\ref{frt}),
the field equations in $f(R,T)$ gravity is given by,
\begin{eqnarray}
G_{\mu \nu}&=&8\pi T_{\mu \nu}^{\text{eff}},
\end{eqnarray}
where $G_{\mu \nu}$ is the Einstein tensor and
\begin{eqnarray}
T_{\mu \nu}^{\text{eff}}&=& T_{\mu \nu}+\frac{\chi}{8\pi}T g_{\mu \nu}+\frac{\chi}{4\pi}(T_{\mu \nu}+p g_{\mu \nu}).
\end{eqnarray}
The field equations in modified gravity can be written as,
\begin{eqnarray}
8\pi\rho^{\text{eff}}&=&\frac{\lambda'}{r}e^{-\lambda}+\frac{1}{r^{2}}(1-e^{-\lambda}),\label{f1}\\
8 \pi p^{\text{eff}}&=& \frac{1}{r^{2}}(e^{-\lambda}-1)+\frac{\nu'}{r}e^{-\lambda},\label{f2} \\
8 \pi p^{\text{eff}}&=&\frac{1}{4}e^{-\lambda}\left[2\nu''+\nu'^2-\lambda'\nu'+\frac{2}{r}(\nu'-\lambda')\right]. \label{f3}
\end{eqnarray}
 for the line element (\ref{line}),
where $\rho^{\text{eff}}$ and $p^{\text{eff}}$ are respectively the density and pressure in Einstein Gravity and
\begin{eqnarray}
\rho^{\text{eff}}&=& \rho+\frac{\chi}{8\pi}(3 \rho-p),\label{r1}\\
p^{\text{eff}}&=& p-\frac{\chi}{8\pi}(\rho-3p),\label{r2}
\end{eqnarray}
the prime indicates differentiation with respect to `r'. Using Eqs. (\ref{f1})-(\ref{f3}), we get,
\begin{eqnarray}\label{con}
\frac{\nu'}{2}(\rho+p)+\frac{dp}{dr}&=&\frac{\chi}{8\pi+2\chi}(p'-\rho').
\end{eqnarray}
In eqn.(\ref{con}), for $\chi =0$ we acquire the conservation equation in Einstein gravity. In next section we shall solve the eqns. (\ref{f1})-(\ref{f3}) to obtain the model of compact star in $f(R,\,T)$ gravity.

\section{Exact Solution of our proposed Model for isotropic Stars}\label{sec3}
In this section we want to obtain a model of compact star by solving the system of eqns.(\ref{f1})-(\ref{f3}). For this purpose we employ well known Buchdahal metric {\em ansatz} \cite{Buchdahl:1959zz} that encompasses almost all the known solutions to the static Einstein equations with a perfect fluid source which is given by,
\begin{eqnarray}\label{elambda}
e^{\lambda}&=&\frac{2(1 + Cr^2)}{2 - Cr^2},
\end{eqnarray}
and,
\begin{eqnarray}\label{enu}
e^{\nu}&=& A \Big[\Big(1 + Cr^2\Big)^{3/2} + B\sqrt{2 - Cr^2} \Big(5 + 2C r^2\Big)\Big]^2,
\end{eqnarray}
where $A,\, B$ and $C$ are constant parameters that can be obtained from the matching condition. Here both $A$ and $B$ are dimensionless and $C$ has dimension km$^{-2}$.\\
In terms of the near horizon physics, the Buchdahl sphere, the limiting stable isotropic stellar structure free of exotic matter, is very crucial. This is because the signal produced by gravitational waves and black hole shadow will resemble that of a black hole because the radius of the Buchdahl sphere is between the horizon and the photon sphere. In order to briefly illustrate the isotropic matter distribution within the compact object filled with perfect fluid, in this paper we develop a stellar model in this  alternative theory of gravity. An interesting fact associated with the Buchdahl sphere is that, the result that the extremal limit for a Buchdahl sphere is over extremal relative to a black hole spacetime \cite{Chakraborty:2022jg}.  Also, Buchdahl sphere has some new exciting universal properties such as escape velocity. These universal properties of the Buchdahl sphere strongly resemble those of the black hole spacetimes \cite{Dadhich:2022yuk}, transcending general relativity. For these astrophysical reasons, we chose such a specific relativistic fluid sphere in this work.

In the study of compact stellar objects, the presence of physical and geometric singularities within the star is regarded as one of the most crucial aspects. We investigate the behavior of both metric potentials to see if singularizes exist. Inside the compact stellar structure, the metric potentials should be singularity-free, positive, monotonically increasing, and regular for physical viability and stability of the model.\\
At the center of the star, $e^{\lambda}=1$ and $e^{\nu}=A[1+5\sqrt{2}B]^2$,
  and their derivatives are given by,
  \begin{eqnarray}
  (e^{\lambda})'&=& \frac{12 C r}{(-2 + C r^2)^2},\\
  (e^{\nu})'&=& \frac{6 A C r \Big\{\big(1 + C r^2\big)^{3/2} + B \sqrt{2 - C r^2} \big(5 + 2 C r^2\big)\Big\} \Big(B -
   2 B C r^2 + f_1(r)\Big)}{\sqrt{2 - C r^2}},
 \end{eqnarray}
where the expression of $f_1(r)$ is given later.\\
  The derivative of the metric coefficients vanishes at the centre of the star, implying that the metric coefficients are regular at the centre of the star. \begin{figure}[htbp]
    \centering
        \includegraphics[scale=.55]{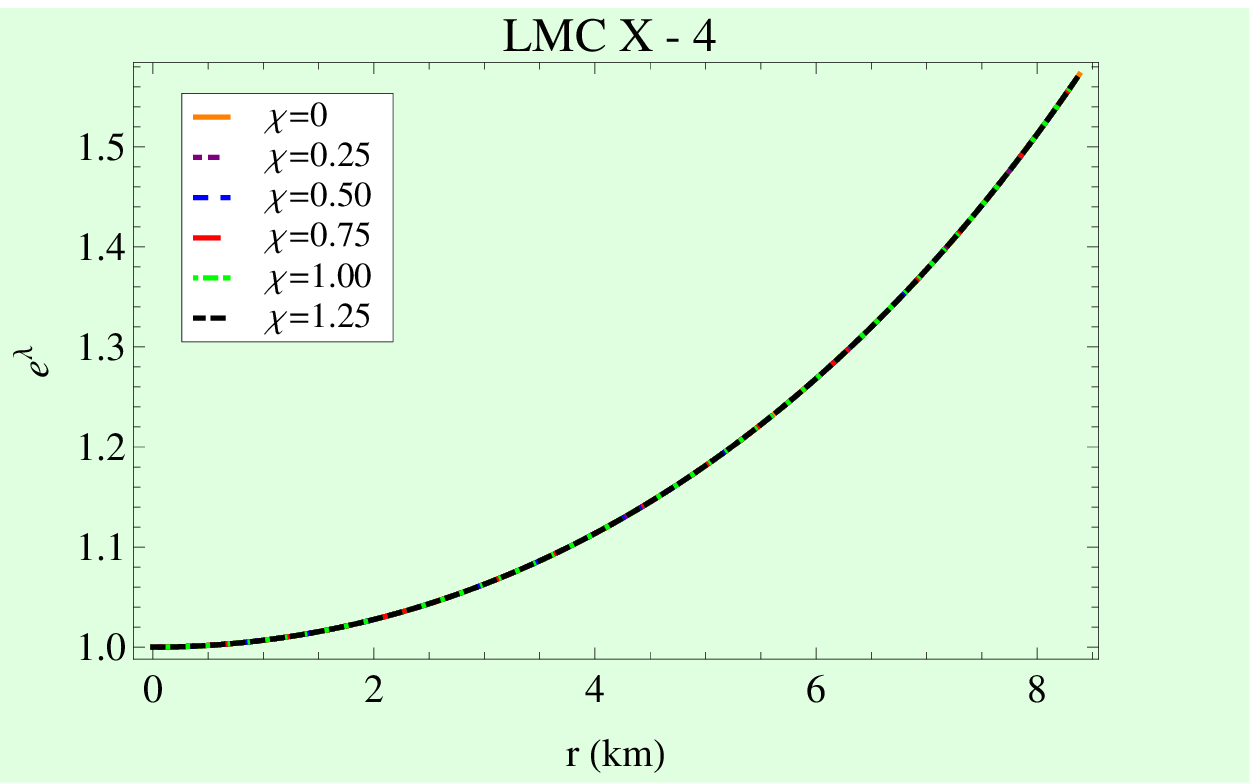}
        \includegraphics[scale=.55]{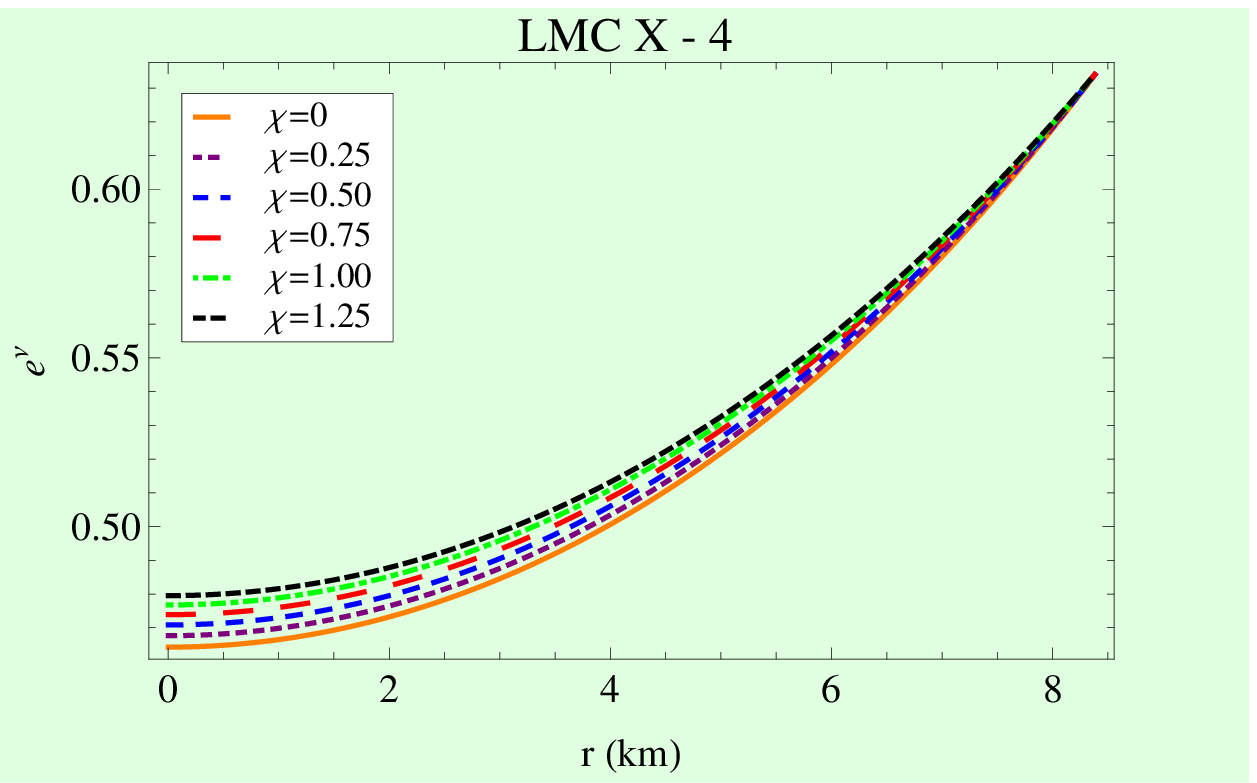}
       \caption{$e^{\lambda}$ and $e^{\nu}$ are shown against `r'.}\label{metric}
\end{figure}
Fig.~\ref{metric} depicts the characteristics of metric coefficients. Both the metric potentials are found to be consistent with the aforementioned conditions. The graphical behavior reveals that the value of both metric potentials is minimum at the centre, then increases nonlinearly until it reaches its maximum at the boundary surface.\par
To explore the entire structure of stellar models with $f(R,\, T)$ gravity, we must first obtain the expressions for physical parameters such as effective pressure and density. The effective density and pressure are calculated by using the expressions of metric potential as,
\begin{eqnarray}
\rho^{\text{eff}}&=&\frac{3 C (3 + C r^2)}{16\pi(1 +  C r^2)^2},\\
p^{\text{eff}}&=&\frac{9 C \Big[B \Big(-2 + C r^2\Big) \Big(1 + 2 C r^2\Big) + \Big(1 - C r^2\Big) f_1(r) \Big]}{16\pi\Big(1 + C r^2\Big)\Big[-B \Big(-2 + C r^2\Big) \Big(5 + 2 C r^2\Big) + \Big(1 + C r^2\Big) f_1(r) \Big]},
\end{eqnarray}
where $f_1(r)= \sqrt{2 + C r^2 - C^2 r^4}$.\\
Using the expression of $p^{\text{eff}}$ and $\rho^{\text{eff}}$, from eqns. (\ref{r1}) and (\ref{r2}), we obtain the expression of matter density and pressure $\rho,\,p$ in modified gravity as,
\begin{eqnarray}
\rho&=&\frac{3 C}{8 \big(\chi + 2 \pi\big) \big(\chi + 4 \pi\big) \big(1 +
   C r^2\big)^2 \Big\{-B \big(-2 + C r^2\big) \big(5 + 2 C r^2\big) + \big(1 + C r^2\big) f_1(r)\Big\}} \Bigg[2 f_1(r) \big(1 + C r^2\big)\times\nonumber\\&&  \Big\{3 \chi + 2 \pi \big(3 + C r^2\big)\Big\} -
   B \big(-2 + C r^2\big) \Big\{4 \pi \big(3 + C r^2\big) \big(5 + 2 C r^2\big) +
      3 \chi \big(7 + 4 C r^2\big)\Big\}\Bigg],\label{p1}\\
      p&=&\frac{3C}{8 \big(\chi + 2 \pi\big) \big(\chi + 4 \pi\big) \big(1 +
   C r^2\big)^2 \Big\{-B \big(-2 + C r^2\big) \big(5 + 2 C r^2\big) + \big(1 + C r^2\big) f_1(r)\Big\}}\Bigg[2 f_1(r)  \Big\{3 \chi + 6 \pi + C \chi r^2 - \nonumber\\ &&
    2 C^2 \big(\chi + 3 \pi\big) r^4\Big\} +
 B \big(-2 + C r^2\big) \Big\{-3 \chi + 12 \pi + 4 C \big(2 \chi + 9 \pi\big) r^2 +
    8 C^2 \big(\chi + 3 \pi\big) r^4\Big\}\Bigg]. \label{p2}
\end{eqnarray}
\begin{figure}[htbp]
    \centering
        \includegraphics[scale=.55]{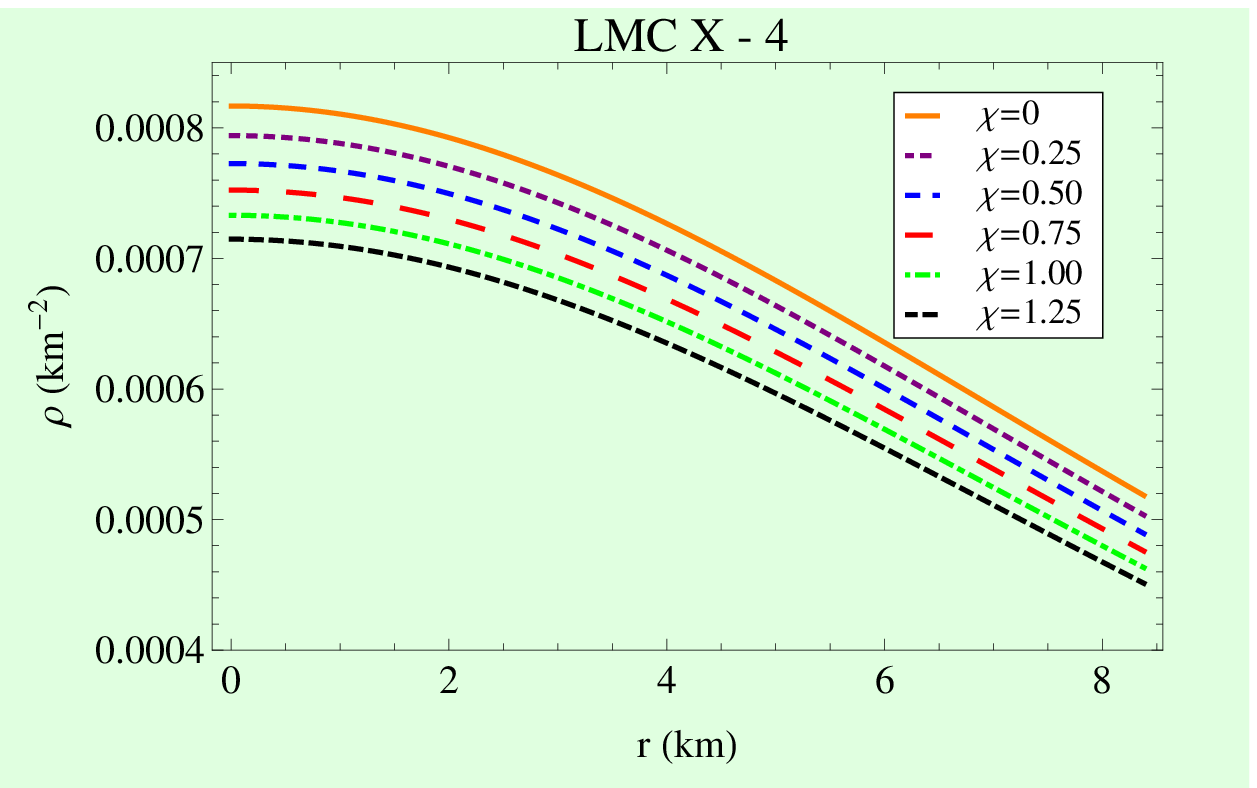}
        \includegraphics[scale=.55]{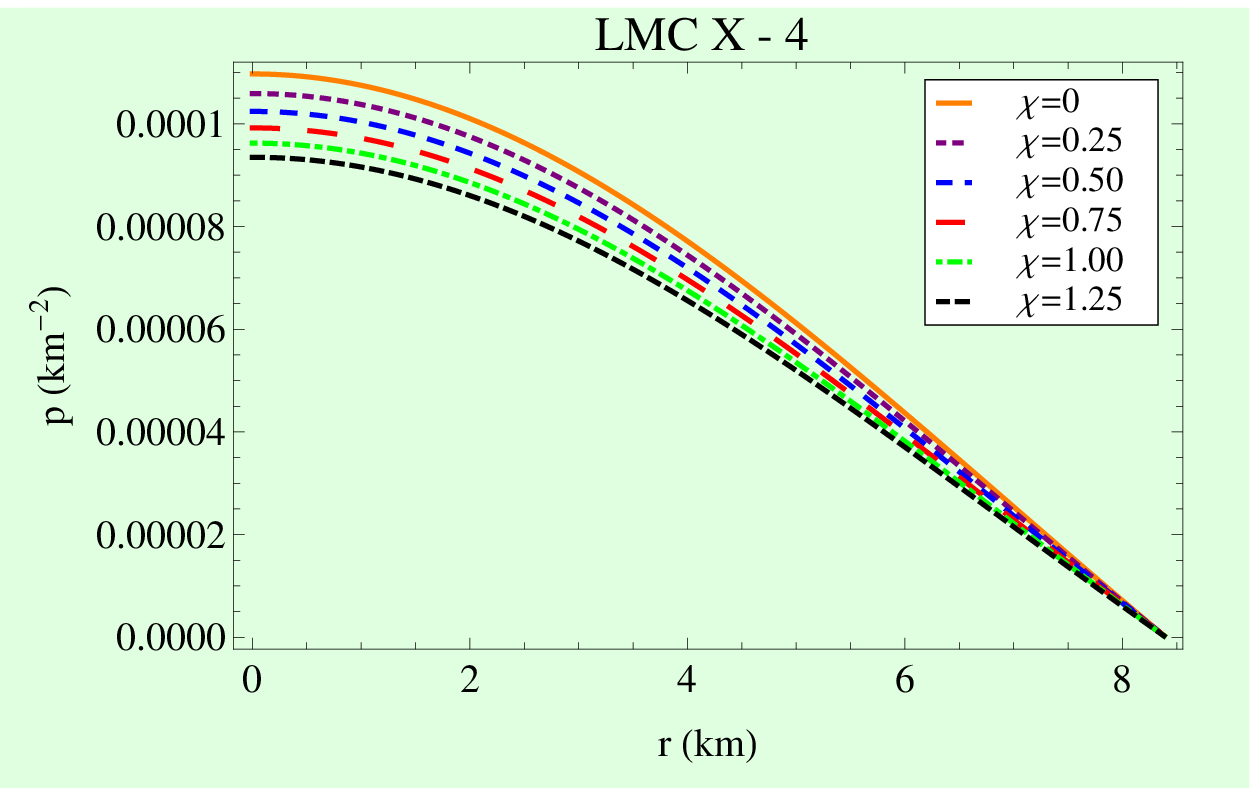}
        \caption{(left) Matter density and (right) pressure are plotted against radius for different values of the coupling constant mentioned in the figure. \label{pp}}
\end{figure}
The profiles of pressure and density are shown in Fig.~\ref{pp} for different values of $\chi$. We can see that the two physical variables are maximum at the origin and decrease monotonously to reach their minimal values at the surface, proving the physical availability of the predicted stellar model. These figures also show that the energy density and pressures at the origin are positive and regular, demonstrating that our framework is free from physical and mathematical singularities.

\section{Exterior line element and matching conditions}\label{sec4}
At the boundary $r=R$, we now match our interior spacetime to the exterior Schwarzschild line element. Corresponding to the interior spacetime,
 \begin{eqnarray}
ds_{-}^2 & = &-A \Big[\Big(1 + Cr^2\Big)^{3/2} + B\sqrt{2 - Cr^2} \Big(5 + 2C r^2\Big)\Big]^2 dt^2+ \frac{2(1 + Cr^2)}{2 - Cr^2} dr^2+r^2(d\theta^2+\sin^2 \theta d\phi^2),
\end{eqnarray}
the exterior line element is described by,
\begin{eqnarray}
ds_+^{2}&=&-\left(1-\frac{2M}{r}\right)dt^{2}+\left(1-\frac{2M}{r}\right)^{-1}dr^{2}+r^{2}\left(d\theta^{2}+\sin^{2}\theta d\phi^{2}\right),
\end{eqnarray}
where `M' denotes the total mass within the boundary of the compact star.\\
The following expressions follow from the continuity of the metric potentials at the boundary surface $r = R$ :
$$g_{rr}^+=g_{rr}^-,\, ~~\text{and} ~~~~g_{tt}^+=g_{tt}^-,$$
where ($-$) and ($+$) sign represent interior and exterior spacetime, respectively. The above two relationships imply,
\begin{eqnarray}
\left(1-\frac{2M}{R}\right)^{-1}&=&\frac{2(1 + CR^2)}{2 - CR^2},\label{o1}\\
1-\frac{2M}{R}&=& A \Big[\Big(1 + CR^2\Big)^{3/2} + B\sqrt{2 - CR^2} \Big(5 + 2C R^2\Big)\Big]^2,\label{o2}
\end{eqnarray}
It is also required that the isotropic pressure `p' vanishes at the boundary `R' i.e., $p(r=R)=0$, implies the following equation:
\begin{align}
\frac{3C}{8 \big(\chi + 2 \pi\big) \big(\chi + 4 \pi\big) \big(1 +
   C r^2\big)^2 \Big\{-B \big(-2 + C r^2\big) \big(5 + 2 C r^2\big) + \big(1 + C r^2\big) f_1(r)\Big\}}\Bigg[2 f_1(r) \Big\{3 \chi + 6 \pi + C \chi r^2 - \nonumber\\
    2 C^2 \big(\chi + 3 \pi\big) r^4\Big\} +
 B \big(-2 + C r^2\big) \Big\{-3 \chi + 12 \pi + 4 C \big(2 \chi + 9 \pi\big) r^2 +
    8 C^2 \big(\chi + 3 \pi\big) r^4\Big\}\Bigg] = 0 .\label{o3}
    \end{align}
 We obtain the expressions for $A,\,B$ and $C$ as follows by solving the equations (\ref{o1})-(\ref{o3}) simultaneously,
  \begin{eqnarray*}
 A &=& \frac{1-\frac{2M}{R}}{3 Q^2},\\
      B &=& -\frac{\sqrt{2} \big(4 M - 3 R\big) R \sqrt{\frac{R (-2 M + R)}{(4 M - 3 R)^2}} \Big\{4 M \big(5 \chi + 12 \pi\big) - 9 \big(\chi + 2 \pi\big) R\Big\}}{(2 M - R) \Big\{-16 \chi M^2 + 8 M \big(7 \chi + 6 \pi\big) R -
   9 \big(\chi - 4 \pi\big) R^2\Big\}},\\
      C &=& -\frac{4 M}{(4 M - 3 R) R^2},
 \end{eqnarray*}
The expression of the constant $Q$ is given by,
\begin{eqnarray*}
Q= 3 \left(\frac{R}{3R-4M}\right)^{3/2}-\frac{6R (4 M - 5 R) \sqrt{\frac{2 M - R}{4 M - 3 R}}  \sqrt{\frac{R (-2 M + R)}{(4 M - 3 R)^2}} \Big\{4 M \big(5 \chi + 12 \pi\big) - 9 \big(\chi + 2 \pi\big) R\Big\}}{(2 M - R) \Big\{-16 \chi M^2 + 8 M \big(7 \chi + 6 \pi\big) R -
   9 \big(\chi - 4 \pi\big) R^2\Big\}}
 \end{eqnarray*}
The approximated mass and radius of the compact star LMC X-4 are used to determine these constant values of $A,\, B$ and $C$ which are listed in Table~\ref{tb12} for different values of $\chi$. It is interesting to note from the table that the numerical values of $A$ is decreasing with the increasing value of $\chi$. On contrary the values of $B$ increases with increasing value of $\chi$ but $C$ does not depend on $\chi$.

\begin{table*}[t]
\centering
\caption{The values of the constants $A,\,B$ and $C$ for the compact star LMC X-4 for different values of coupling constant $\chi$.}
\label{tb12}
\begin{tabular}{@{}cccccccccccccccc@{}}
\hline
Objects  & Estimated &Estimated & $\chi$ & $A$& $B$  &$C$ \\
&Mass ($M_{\odot}$)& Radius &&&  &(km$^{-2}$)\\
\hline
LMC X-4  \cite{Rawls:2011jw}& $1.04$&$8.4$& 0& $0.0360397$&$0.366157$&$0.00456155$\\
&&& $0.25$& $0.0340651$&$0.38256$&$0.00456155$                                    \\
&&& $0.5$ &  $0.0322527$&$0.398922$&$0.00456155$\\
&&& $0.75$ &  $0.0305852$&$0.415244$&$0.00456155$\\
&&& $1$ & $0.0290474$&$0.431526$&$0.00456155$\\
&&& $1.25$ & $0.027626$&$0.447768$&$0.00456155$\\
\hline
\end{tabular}
\end{table*}

\section{Physical properties of the astrophysical structure in $f(R,\,T)$ gravity theory}\label{sec5}
In this section, we will test physical highlights of the stellar structure in the context of $f(R,\,T)$ theory in order to investigate the Modified TOV equation, energy conditions, the status of the sound speed within the stellar system, compactness and gravitational surface redshift, the adiabic index, and so on for different values of the coupling constant $\chi$.
\subsection{Nature of pressure and density}
To check the non singularity behavior of the pressure and density, we calculate the central pressure and central density as,
\begin{eqnarray}
\rho_c &=&\frac{9 C \Big\{\big(2 + 7 \sqrt{2} B\big) \chi + 4 \big(\pi + 5 \sqrt{2} B \pi\big)\Big\}}{8 \big(1 + 5 \sqrt{2} B\big) \big(\chi + 2 \pi\big) \big(\chi + 4 \pi\big)},\\
 p_c&=&\frac{9 C \big(2 \chi + \sqrt{2} B \chi + 4 \pi - 4 \sqrt{2} B \pi\big)}{8 \big(1 + 5 \sqrt{2} B\big) \big(\chi + 2 \pi\big) \big(\chi + 4 \pi\big)}.
\end{eqnarray}
Clearly both $\rho_c$ and $p_c$ are finite.\\
The pressure and density gradient can be obtained by taking the differentiation of the expressions of $\rho$ and $p$ given in eqns.(\ref{p1})-(\ref{p2}), which yields,
\begin{eqnarray*}
\rho'&=& \frac{f_2(r)}{f_3(r)},\\
p'&=& \frac{f_4(r)}{f_3(r)}.
\end{eqnarray*}
where the expressions of $f_2,\,f_3$ and $f_4$ are given by,
\begin{eqnarray*}
f_2(r) &=& 3 C^2 r \Bigg[-8 \big(1 + C r^2\big)^3 C_1 \Big\{3 \chi + \pi \big(5 + C r^2\big)\Big\} +
   B (1 + C r^2)^2 \Big\{16 \pi \big(-2 + C r^2\big) \big(5 + C r^2\big) \big(5 + 2 C r^2\big) \nonumber\\ && +
      \chi \big(-345 - 6 C r^2 + 96 C^2 r^4\big)\Big\} +
   4 B^2 \big(-2 + C r^2\big) C_1 \Bigg\{2 \pi \big(5 + C r^2\big) \big(5 + 2 C r^2\big)^2  +
      3 \chi \bigg(32 + \nonumber\\ && C r^2 \big(31 + 8 C r^2\big)\bigg)\Bigg\}\Bigg],\nonumber\\
f_3(r) &=& 8 \big(\chi + 2 \pi\big) \big(\chi + 4 \pi\big) \sqrt{2 - C r^2} \big(1 + C r^2\big)^{7/2} \Big\{\big(1 + C r^2\big)^{3/2} + B \sqrt{2 - C r^2} \big(5 + 2 C r^2\big)\Big\}^2,\nonumber\\
f_4(r) &=& -\Bigg[3 C^2 r \Bigg\{-8 \big(1 + C r^2\big)^3 \Big\{-4 \chi - 9 \pi +
        c \big(\chi + 3 \pi\big) r^2\Big\} C_1 +
     4 B^2 \big(-2 + C r^2\big) C_1  \Big\{-38 \chi - 18 \pi + \nonumber\\ && C \big(-29 \chi + 6 \pi\big) r^2 +
        8 C^2 \big(\chi + 6 \pi\big) r^4 + 8 C^3 \big(\chi + 3 \pi\big) r^6\Big\} +
     B \big(1 + C r^2\big)^2 \Big\{235 \chi + 360 \pi -
        2 C \big(31 \chi + 96 \pi\big) r^2 \nonumber\\ &&-
        16 C^2 \big(7 \chi + 15 \pi\big) r^4 +
        32 C^3 \big(\chi + 3 \pi\big) r^6\Big\}\Bigg\}\Bigg].
\end{eqnarray*}
\begin{figure}[htbp]
    \centering
        \includegraphics[scale=.55]{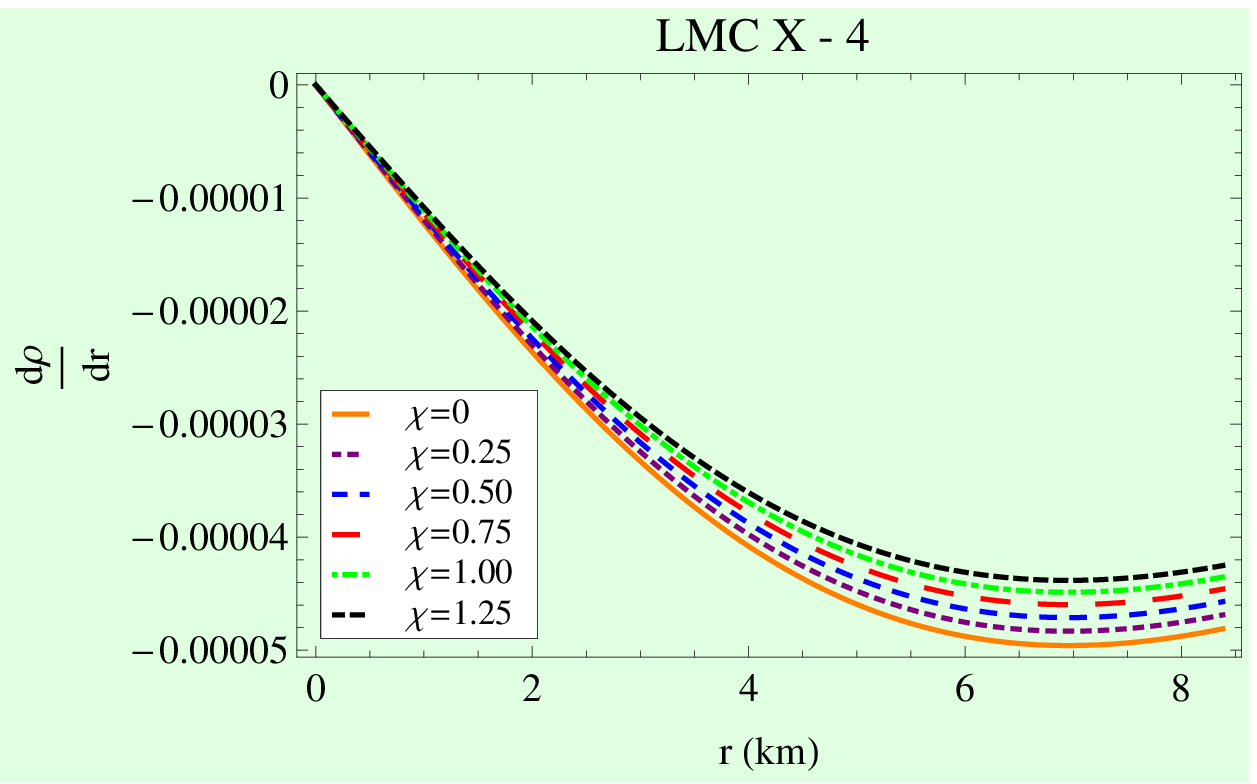}
        \includegraphics[scale=.55]{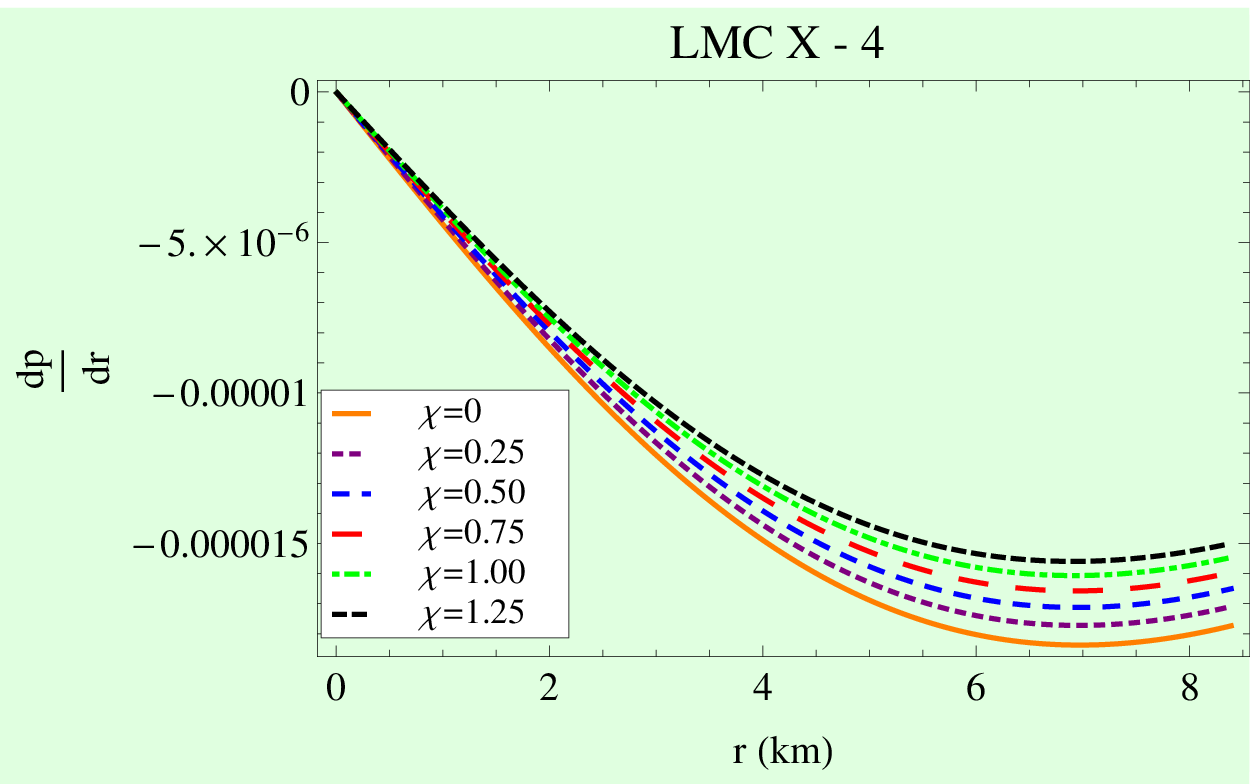}
       \caption{The pressure and density gradients are shown against `r'.}\label{grad5}
\end{figure}
\begin{figure}[htbp]
    \centering
        \includegraphics[scale=.55]{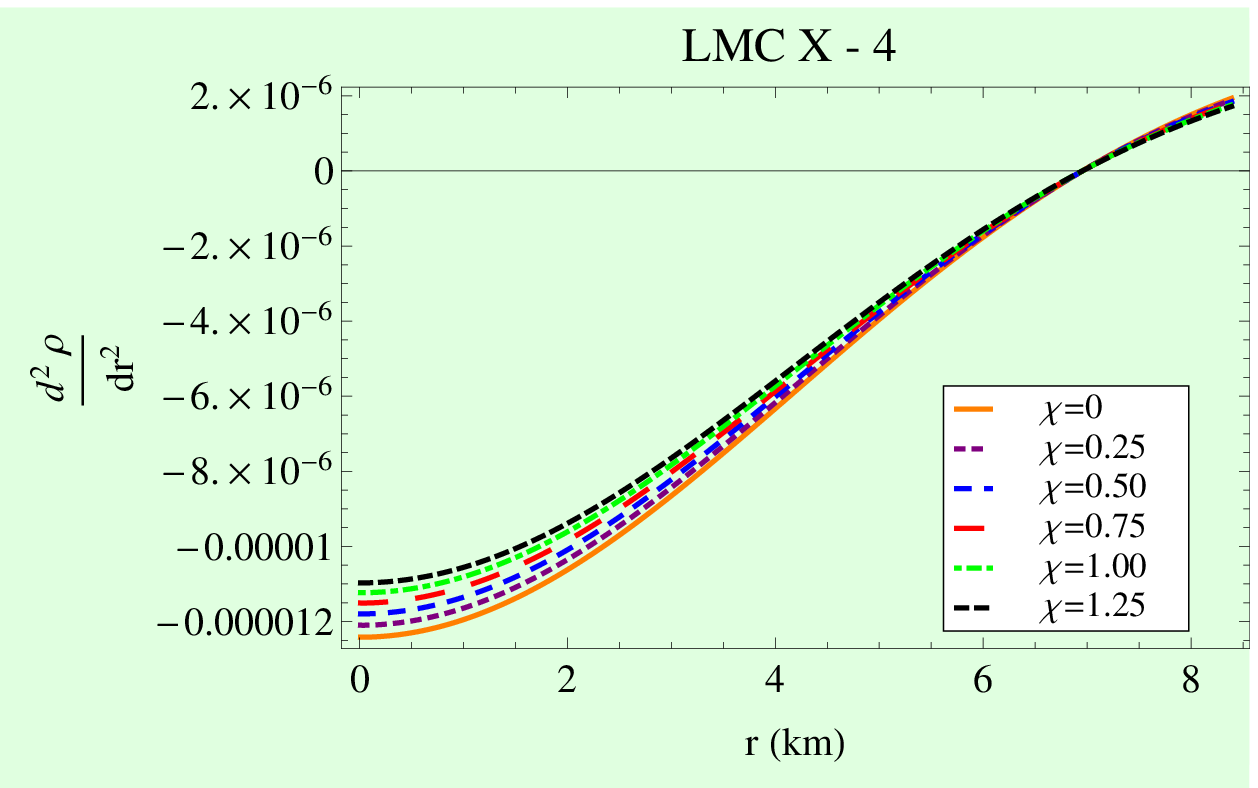}
        \includegraphics[scale=.55]{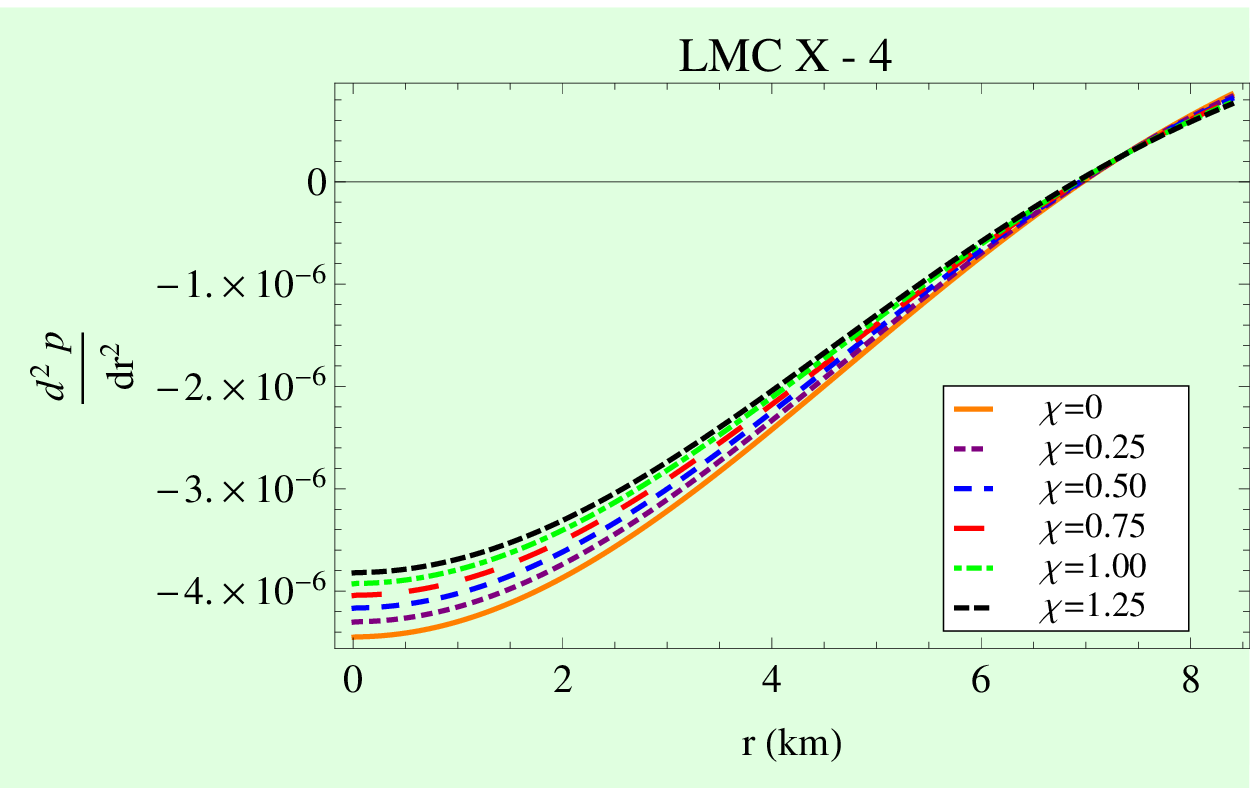}
       \caption{$\frac{d^2\rho}{dr^2}$ and $\frac{d^2 p}{dr^2}$ are shown against `r'.}\label{rho2}
\end{figure}

The behavior of pressure and density gradient are shown in Fig.~\ref{grad5} for different values of $\chi$. From the figures one can note that $\rho',\,p'<0$ in the interior of the stellar model and $\rho'(0)=0=p'(0)$. Moreover at the center of the star $\rho''(0),\,p''(0)<0$ as shown in Fig.~\ref{rho2}.\par

 \subsection{Energy Conditions}
 Energy conditions are a set of physical properties that can be used to explore the presence of ordinary and exotic matter inside a star formation. The validity of the second law of black hole thermodynamics and the Hawking-Penrose singularity theorems can be easily tested using the energy conditions \cite{Hawking:1973uf}. These conditions of energy are referred to as null, weak, strong and dominant energy conditions, symbolized respectively by NEC, WEC, SEC and DEC. All energy conditions for our current model are met if the following inequalities are hold :
    \begin{eqnarray*}
 \text{NEC}:~\rho+p\geq 0,\, \text{WEC}:~\rho+p\geq 0,~ \rho \geq 0,\, \text{SEC}:~\rho+p \geq 0, \rho+ 3p \geq 0,\,\text{DEC}:~\rho-p\geq 0,~ \rho \geq 0.
\end{eqnarray*}
To check the aforementioned energy conditions, we shall require the following expressions.
\begin{eqnarray}
\rho+p&=&\frac{3 C \left[B (-2 + C r^2)^2 (3 + 2 C r^2) +
   \sqrt{2 + C r^2 - C^2 r^4} \Big(3 + C r^2 (2 - C r^2)\Big)\right]}{2 (\chi + 4 \pi) (1 +
   Cr^2)^2 \left[-B (-2 + C r^2) (5 + 2 C r^2) + (1 + C r^2) \sqrt{
    2 + C r^2 - C^2 r^4}\right]},\\
\rho+3p&=&\frac{3 C }{4 (\gamma + 2 \pi) (\gamma + 4 \pi) (1 +
   C r^2)^2 h_1(r)}\times\left[-2 (1 + C r^2) \left(-6 (\chi+ 2 \pi) +
      C (3\chi + 8 \pi) r^2\right) \sqrt{2 + C r^2 - C^2 r^4} \right.\nonumber\\&&\left.+
   B (-2 + Cr^2) \left\{3 \chi (-5 + 2 C r^2 + 4 C^2 r^4) +
      4 \pi \left(-3 + 8 C r^2 (1 + C r^2)\right)\right\}\right],\\
   \rho-p&=&\frac{3 C \Big\{2 C r^2 (1 + Cr^2) \sqrt{2 + C r^2 - C^2 r^4} -
   B (-2 + C r^2) (9 + 2 C r^2 (5 + 2 C r^2))\Big\}}{4 (\gamma + 2 \pi) (1 + Cr^2)^2 h_1(r)},
\end{eqnarray}
where the expression of $h_1(r)$ is given by,
\[h_1(r)=-B (-2 + Cr^2) (5 + 2 Cr^2) + (1 + Cr^2) \sqrt{
    2 + c r^2 - c^2 r^4}.\]

 \begin{figure}[htbp]
    \centering
        \includegraphics[scale=.39]{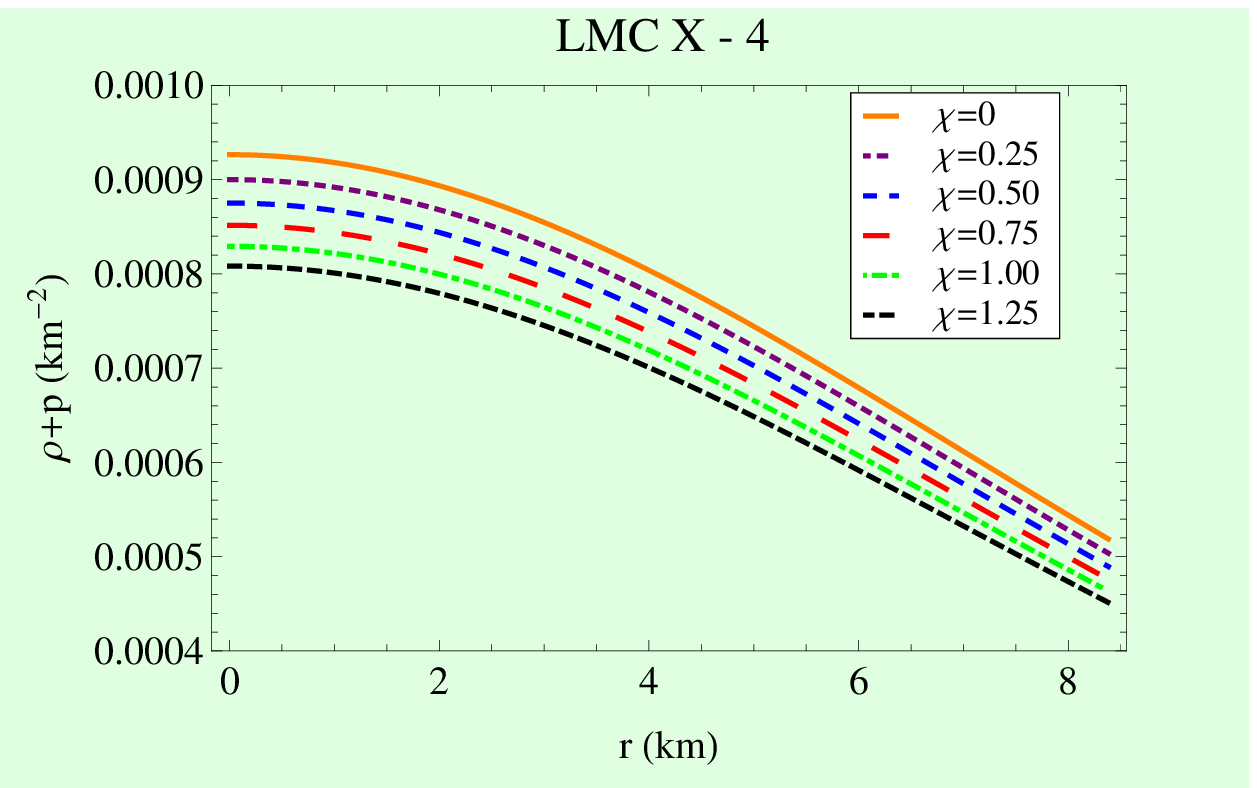}
        \includegraphics[scale=.39]{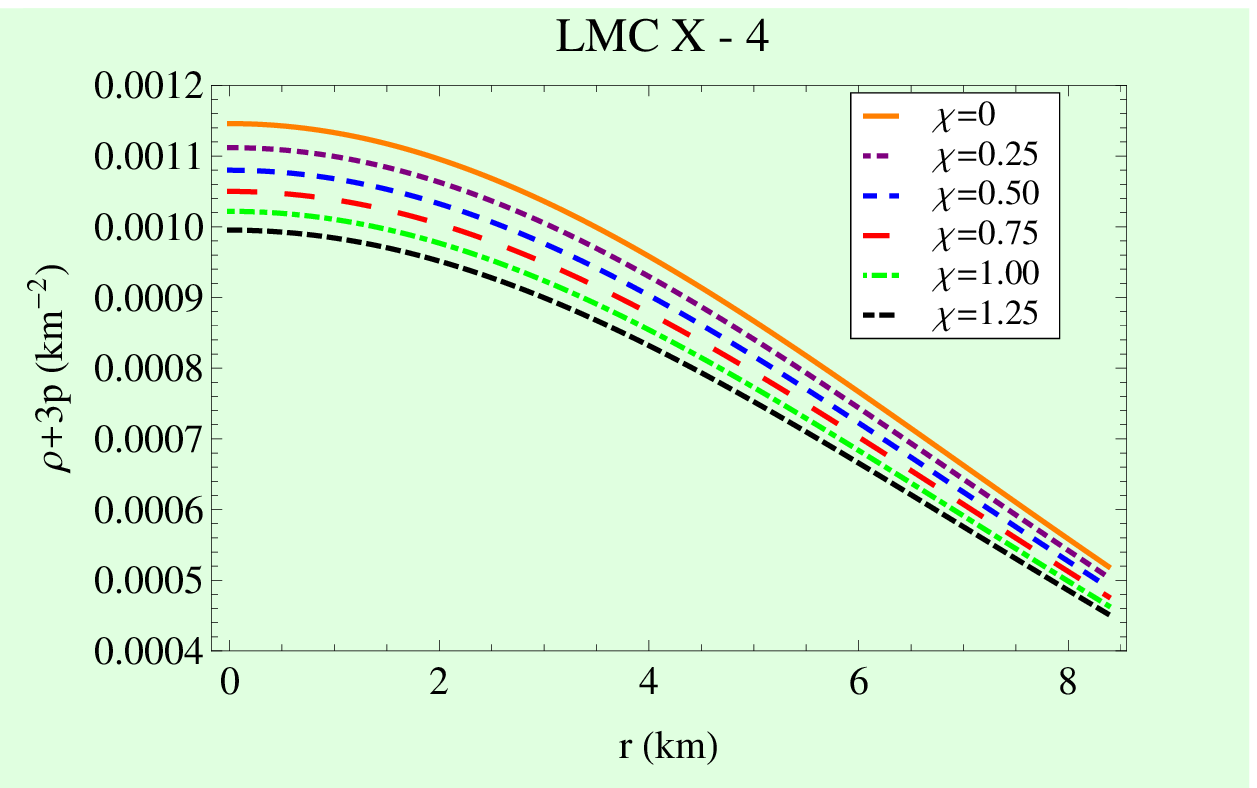}
        \includegraphics[scale=.39]{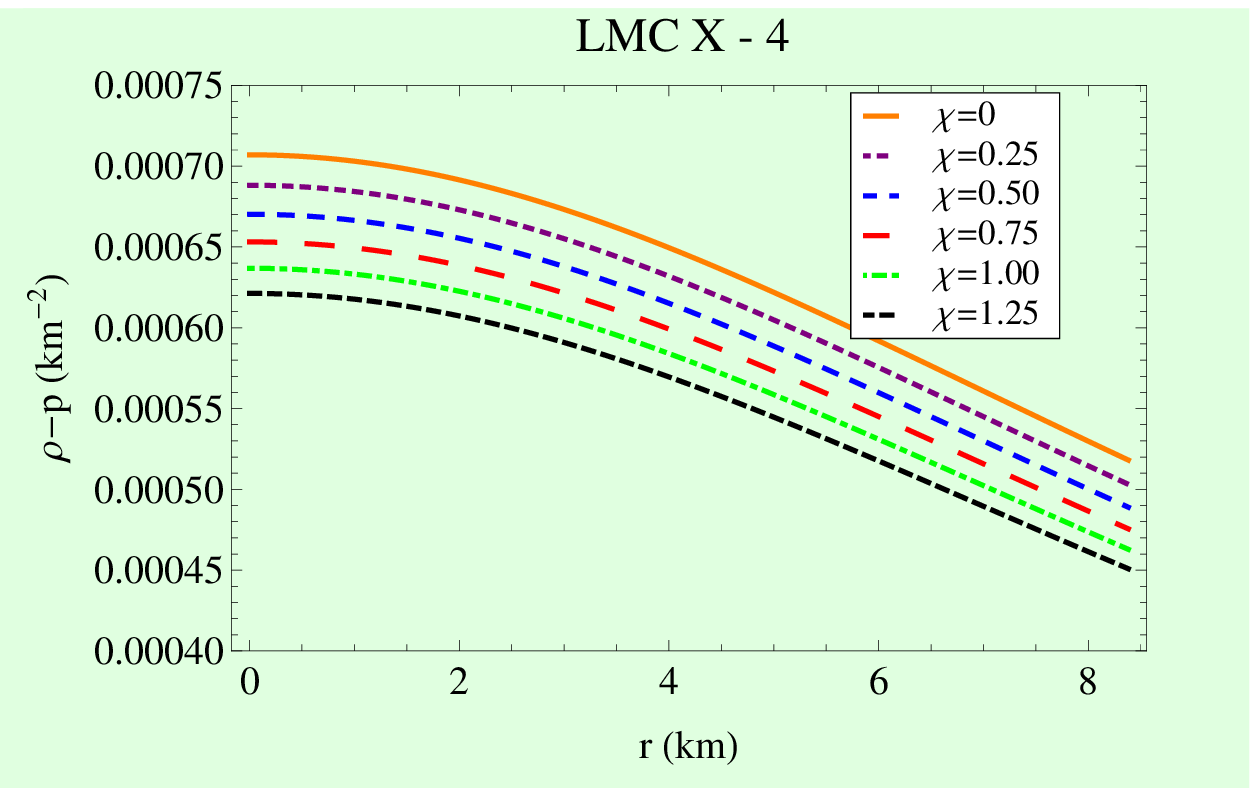}
       \caption{The energy conditions are shown against `r'.}\label{ener1}
\end{figure}

All of the energy conditions for our chosen $f(R,\,T)$ model have been met, as shown graphically in Fig.~\ref{ener1}.
\subsection{Causality condition}
Now we investigate the stability criteria for a physically realistic anisotropic stellar compact object using graphical representation by employing numerical values for several unknown constants. We use the causality condition to demonstrate this criteria. The square of the speed of sound $V^2$ in the entire region of the fluid sphere must follow the bound $0<V^2<1$ to satisfy causality condition.
\begin{eqnarray}
V^2&=&\frac{h_2(r)}{h_3(r)},
\end{eqnarray}
where,
\begin{eqnarray*}
h_2(r) &=& 8 \big(1 + C r^2\big)^3 \Big\{-4 \chi - 9 \pi + C \big(\chi + 3 \pi\big) r^2\Big\} C_1 -
 4 B^2 \big(-2 + C r^2\big) C_1 \Big\{-38 \chi
 - 18 \pi + C \big(-29 \chi + 6 \pi\big) r^2 \\&& + 8 C^2 \big(\chi + 6 \pi\big) r^4 + 8 C^3 \big(\chi + 3 \pi\big) r^6\Big\} - B \big(1 + C r^2\big)^2 \Big\{235 \chi + 360 \pi - 2 C \big(31 \chi  + 96 \pi\big) r^2 - 16 C^2 \big(7 \chi \\&& + 15 \pi\big) r^4 + 32 C^3 \big(\chi + 3 \pi\big) r^6\Big\}  ,\\
h_3(r) &=& -8 \big(1 + C r^2\big)^3 C_1 \Big\{3 \chi + \pi \big(5 + C r^2\big)\Big\} +
 B \big(1 + C r^2\big)^2 \Big\{16 \pi \big(-2 + C r^2\big) \big(5 + C r^2\big) \big(5 + 2 C r^2\big)
 + \chi (-345 \\&& - 6 C r^2 + 96 C^2 r^4)\Big\} +
 4 B^2 \big(-2 + C r^2\big) C_1 \bigg[2 \pi \big(5 + C r^2\big) \big(5 + 2 C r^2\big)^2 +
    3 \chi \Big\{32 +  C r^2 \big(31 + 8 C r^2\big)\Big\}\bigg]
\end{eqnarray*}
\begin{figure}[htbp]
    \centering
        \includegraphics[scale=.55]{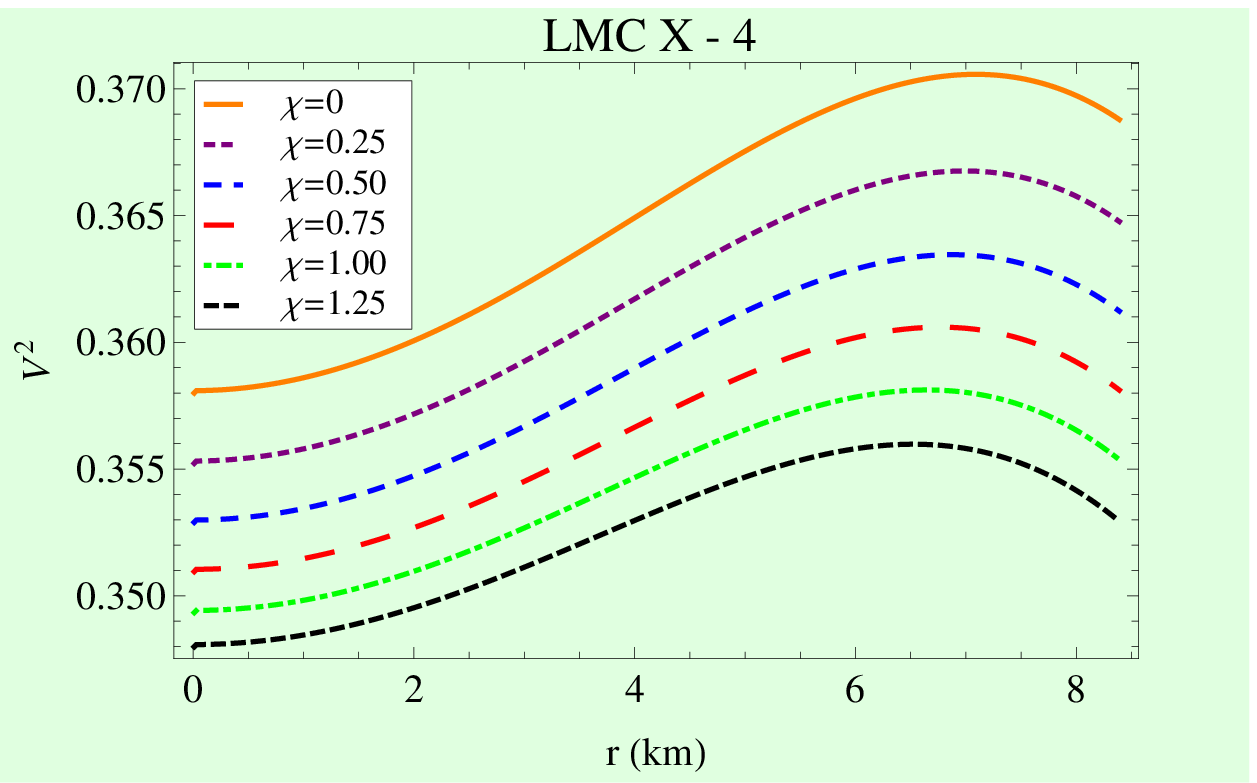}
        \includegraphics[scale=.55]{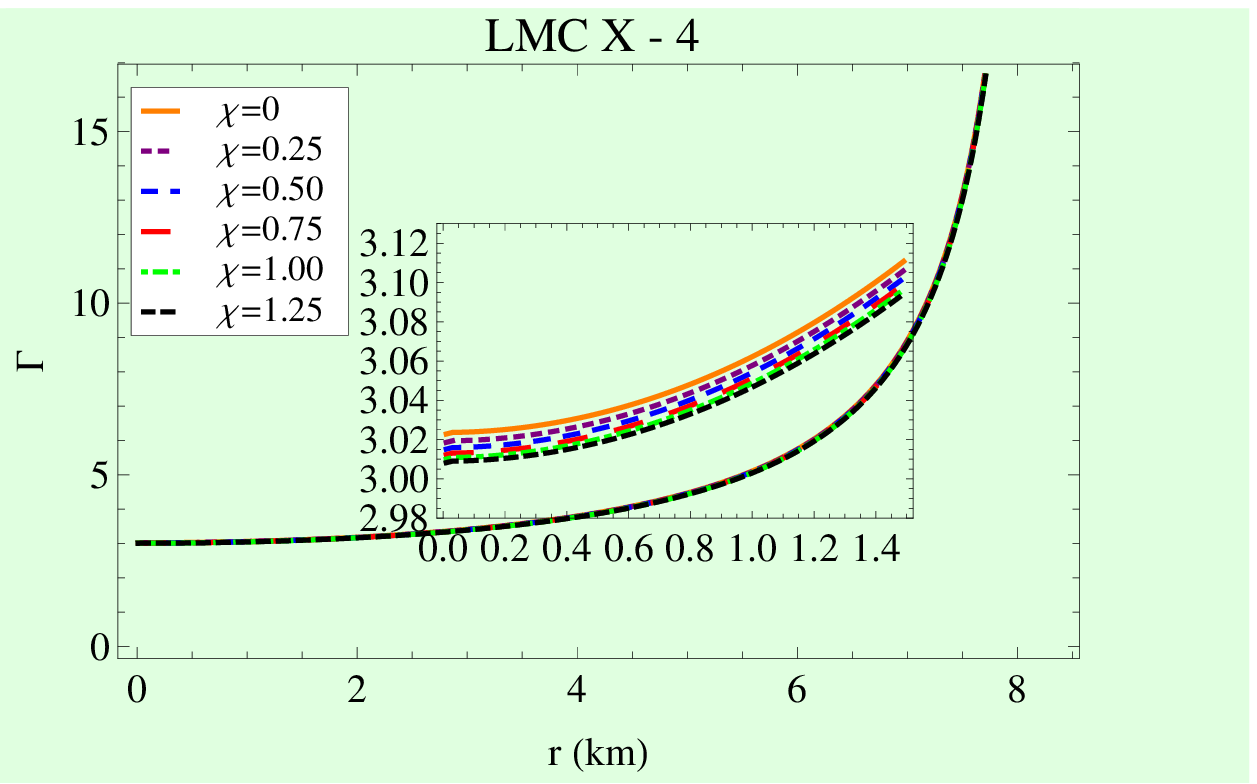}
       \caption{(left) The square of the sound velocity and (right) relativistic adiabatic index are plotted against radius inside the stellar interior.\label{sv}}
\end{figure}
In Fig.~\ref{sv}, the graphical nature of the causality condition for the compact object LMC X-4 is examined for different values of $\chi$, from which it is clear that the square of the sound speed lies within the predicted range throughout the fluid sphere.
\subsection{Relativistic adiabatic index}
For a given energy density, the stiffness of the equation of state can be characterized by the term adiabatic index, which also displays the stability of both relativistic and non-relativistic compact stars. Chandrasekhar \cite{Chandrasekhar:1964zz} introduced the concept of dynamical stability against infinitesimal radial adiabatic perturbation of the stellar system, and previous researches \cite{heintzmann1975neutron,hillebrandt1976anisotropic} have successfully proven this hypothesis for both isotropic and anisotropic stellar objects. According to their estimations, the adiabatic index in all internal points of a dynamically stable stellar object must be greater than $4/3$. The expression of relativistic adiabatic index $\Gamma$ is given by,
\begin{eqnarray}
\Gamma&=&\frac{\rho+p}{p}V^2,\nonumber\\&=&\frac{h_4(r)}{h_5(r)}V^2
\end{eqnarray}
where,
\begin{eqnarray*}
h_4(r) &=& 4 \big(\chi + 2 \pi\big) \Big[B \big(-2 + C r^2\big)^2 \big(3 + 2 C r^2\big) +
   C_1 \Big\{3 + C r^2 \big(2 - C r^2\big)\Big\}\Big],\\
h_5(r) &=& 2 f_1(r) \Big\{3 \chi + 6 \pi + C \chi r^2 -
    2 C^2 \big(\chi + 3 \pi\big) r^4\Big\} +
 B \big(-2 + C r^2\big) \Big\{-3 \chi + 12 \pi + 4 C \big(2 \chi + 9 \pi\big) r^2 \\&&
 + 8 C^2 \big(\chi + 3 \pi\big) r^4\Big\}.
\end{eqnarray*}
Fig.~\ref{sv} depicts the behavior of the adiabatic index $\Gamma$. The value of the adiabatic index is greater than $4/3$, as seen from the graph, confirming the stability of our proposed model.
\subsection{TOV Equation}
The hydrostatic equilibrium equation is an important attribute of the presented physical realistic compact object. By using the generalized Tolman-Oppenheimer-Volkov (TOV) equation, we can evaluate this equilibrium equation for our compact star candidate under the combined behavior of different forces.
\begin{eqnarray}\label{con1}
-\frac{\nu'}{2}(\rho+p)-\frac{dp}{dr}+\frac{\chi}{8\pi+2\chi}(p'-\rho')=0,
\end{eqnarray}
Under the combined action of three different forces, namely gravitational ($F_g$), hydrostatic ($F_h$), and the additional force due to modified gravity ($F_m$), the above equation predicts the stable configuration for the anisotropic celestial compact object.\\
\begin{figure}[htbp]
    \centering
        \includegraphics[scale=.55]{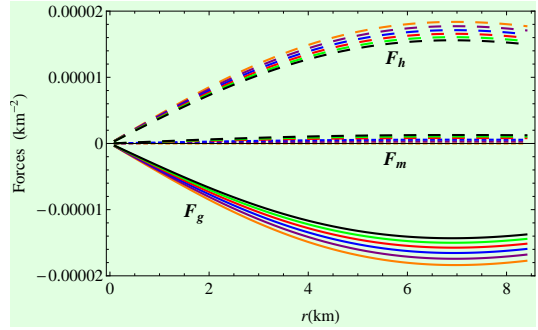}
       \caption{Different forces acting on the system are plotted against radius inside the stellar interior for different values of $\chi$. The color description of the curves are same as Fig.~\ref{metric} \label{tov1}}
\end{figure}
The equation (\ref{con1}) can be written as,
\[F_g+F_h+F_m=0,\]
where,
\begin{eqnarray}
F_g&=&-\frac{9 C^2 r \Big\{B - 2 B C r^2 + C_1\Big\} \Big[B \big(-2 + C r^2\big)^2 \big(3 + 2 C r^2\big) +
   C_1 \Big\{3 + C r^2 \big(2 - C r^2\big)\Big\}\Big]}{2 \big(\chi + 4 \pi\big) \big(-2 + C r^2\big) \big(1 + C r^2\big)^2 \Big[\big(1 + C r^2\big)^{3/2} +
   B \sqrt{2 - C r^2} \big(5 + 2 C r^2\big)\Big]^2},\\
F_h &=&\frac{h_6(r)}{h_7(r)},\\
   F_m &=& -\frac{h_8(r)}{h_7(r)}.
   \end{eqnarray}
   where,
   \begin{eqnarray*}
   h_6(r) &=& 3 C^2 r \Bigg[-8 (1 + C r^2)^3 \Big\{-4 \chi - 9 \pi +
      C (\chi + 3 \pi) r^2\Big\} C_1 +
   4 B^2 (-2 + C r^2) C_1 \Big\{-38 \chi - 18 \pi + C (-29 \chi + 6 \pi) r^2 + \\&&
      8 C^2 (\chi + 6 \pi) r^4 + 8 C^3 (\chi + 3 \pi) r^6\Big\} +
   B (1 + C r^2)^2 \Big\{235 \chi + 360 \pi -
      2 C (31 \chi + 96 \pi) r^2 -
      16 C^2 (7 \chi + 15 \pi) r^4 \\&& + 32 C^3 (\chi + 3 \pi) r^6\Big\}\Bigg] ,\\
   h_7(r) &=& 8 (\chi + 2 \pi) (\chi + 4 \pi) \sqrt{2 - C r^2} (1 + C r^2)^{7/2} \Big\{(1 + C r^2)^{3/2} + B \sqrt{2 - C r^2} (5 + 2 C r^2)\Big\}^2,
   \end{eqnarray*}
   \begin{eqnarray*}
   h_8(r) &=&  3 C^2 \chi r \Bigg[-4 (-1 + C r^2) (1 + C r^2)^3 C_1 +
   4 B^2 C_1 (-58 - 35 C r^2 + 8 C^3 r^6 + 4 C^4 r^8) +
   B (1 + C r^2)^2 \Big\{-55 \\&& + 2 C r^2 (-17 - 4 C r^2 + 8 C^2 r^4)\Big\}\Bigg].
   \end{eqnarray*}
   The profiles of all the forces involved in the hydrostatic equilibrium condition are shown in Fig.~\ref{tov1}. The figure shows that the gravitational force counterbalances the combined behavior of hydrostatic force and modified gravity force, keeping our present system in stable equilibrium.
\subsection{Equation of state} 
\begin{figure}[htbp]
    \centering
        \includegraphics[scale=.55]{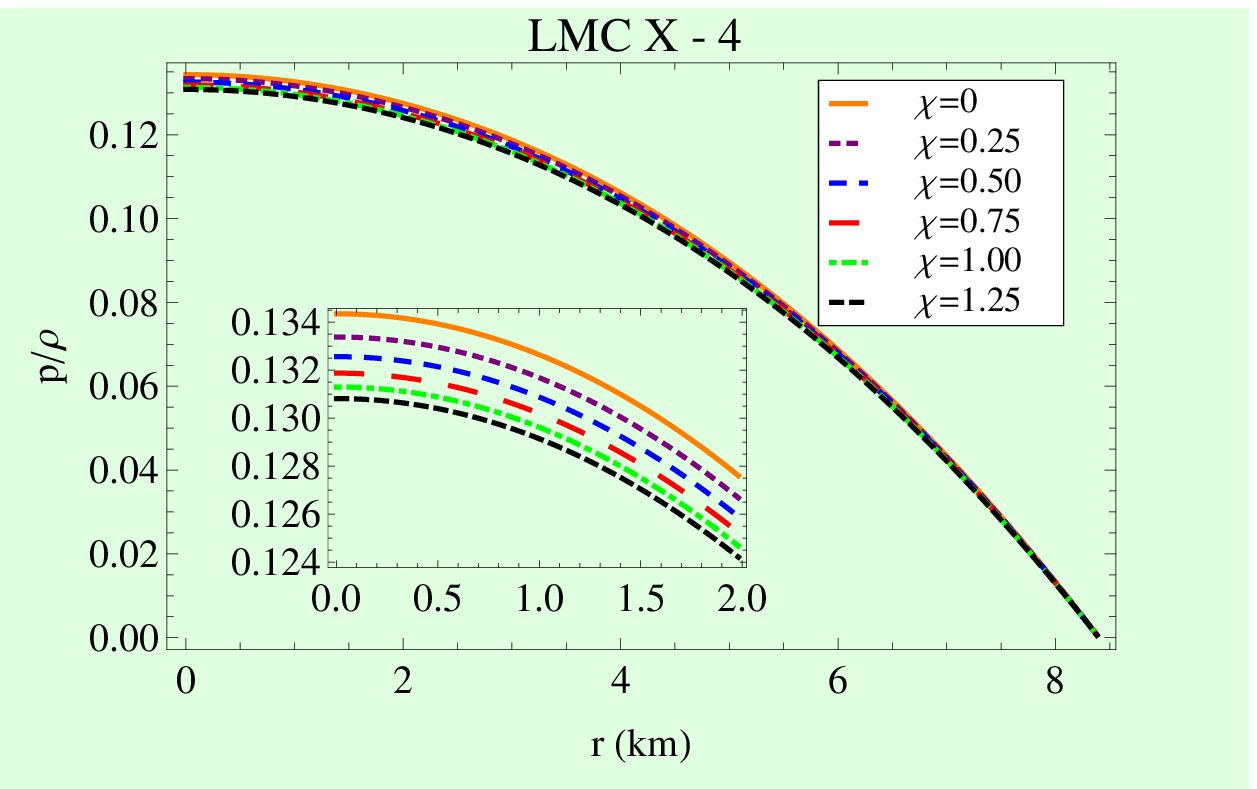}
        \includegraphics[scale=.55]{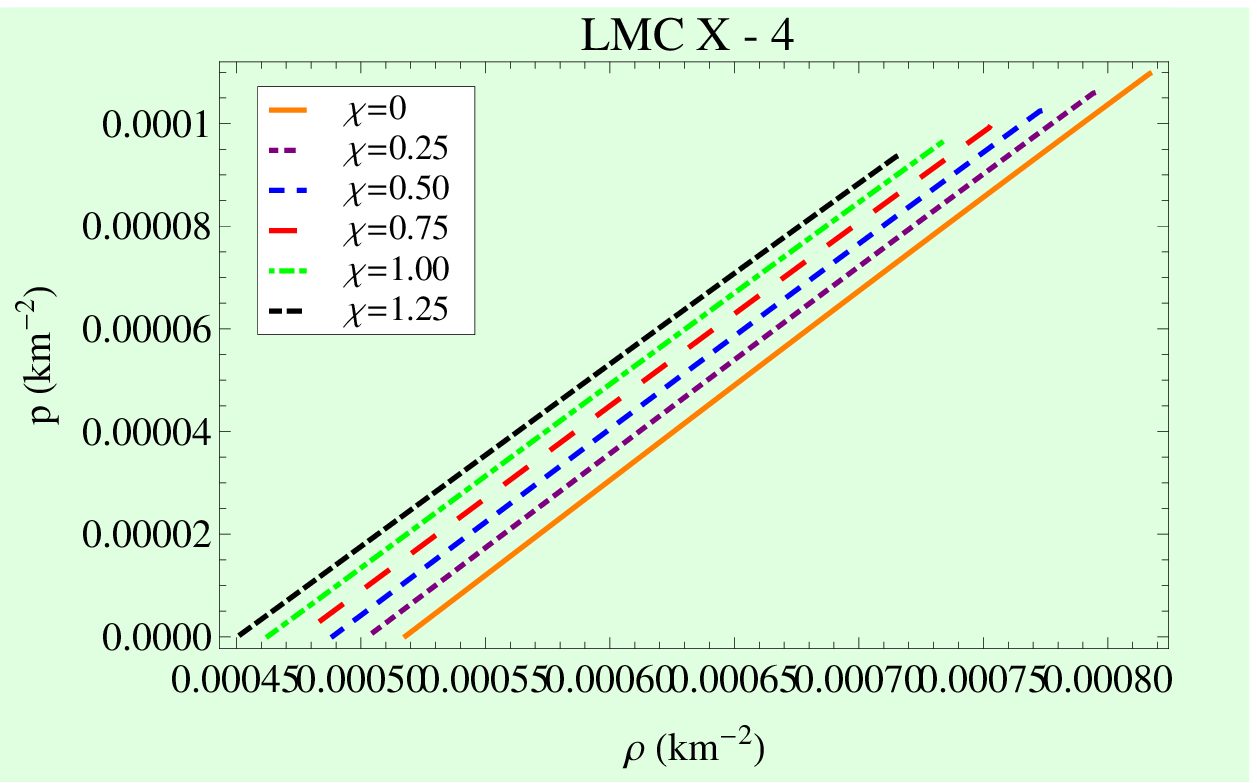}
       \caption{The pressure and density relation are shown inside the stellar interior.}\label{eoss}
\end{figure}

The equation of state describes the connection between pressure and density of matter. To model the compact object, many researchers employed linear, quadratic, polytropic, and other equations of state. To develop the stellar model in this study, we did not assume any specific equation of state. In Fig.~\ref{eoss}, we have depicted the variation of pressure with respect to density using a graphical representation. The ratio of pressure to the density is also depicted in Fig.~\ref{eoss} for different values of $\chi$.
\subsection{Mass radius relationship}
\begin{figure}[htbp]
    \centering
        \includegraphics[scale=.45]{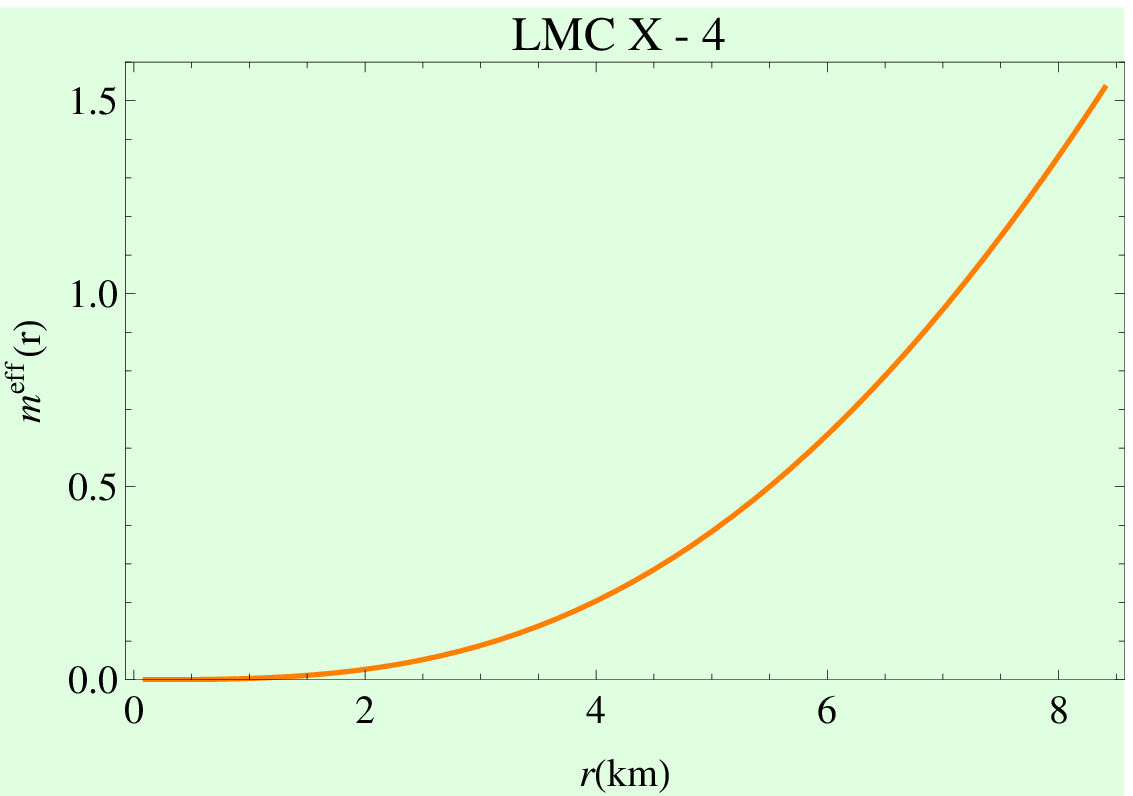}
        \includegraphics[scale=.45]{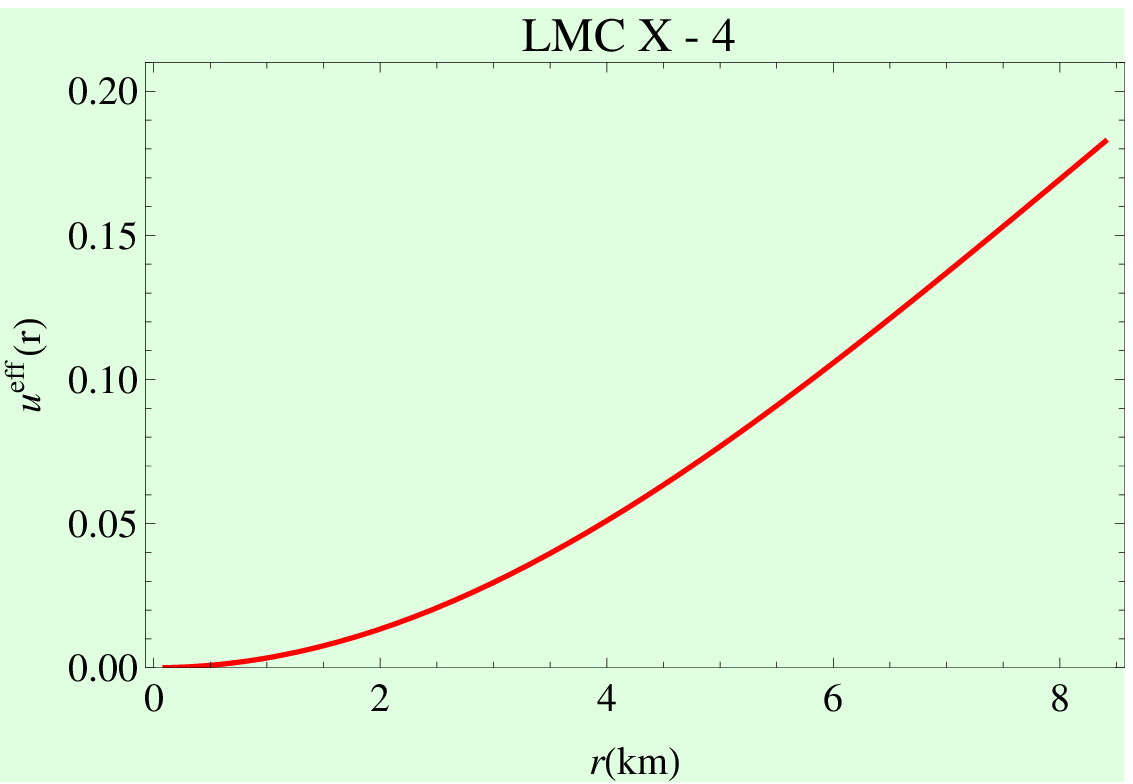}
        \includegraphics[scale=.45]{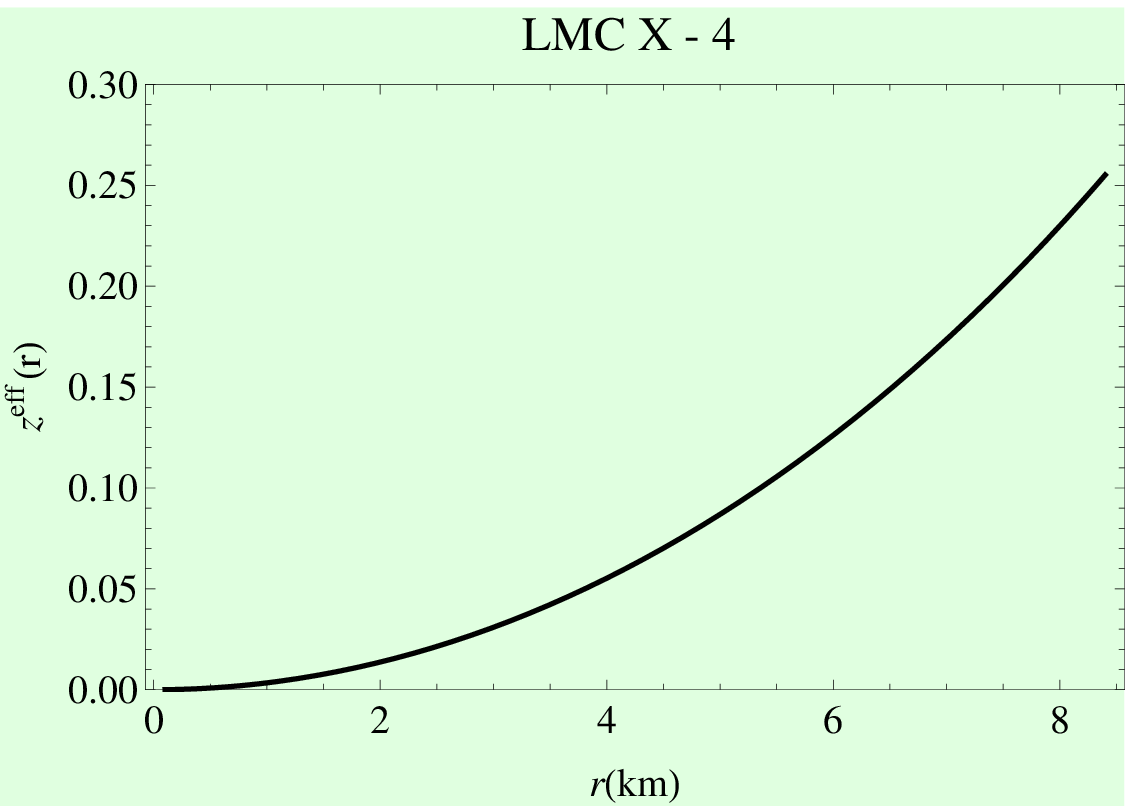}
       \caption{(left) The effective mass (middle) the effective compactness and (right) the effective surface redshift are plotted against radius inside the stellar interior.\label{m11}}
\end{figure}

The mass function $m(r)$ of the present stellar system is determined by,
\begin{eqnarray}
m(r)&=&4\pi\int_0^r \rho r^2 dr,\nonumber\\
&=&\frac{\pi (3 \chi + 8 \pi)}{2(\chi + 2 \pi) (\chi + 4 \pi)}\frac{3 C r^3}{2 (1 + Cr^2)}\nonumber\\&&+\frac{\pi \chi}{2(\chi + 2 \pi) (\chi + 4\pi)}\int_0^r\frac{9C r^2 \left\{B (-2 + C r^2) (1 + 2 C r^2) + (1 - C r^2) \sqrt{
    2 + c r^2 - c^2 r^4}\right\}}{2 (1 +
   C r^2) \left\{-B (-2 + Cr^2) (5 + 2 Cr^2) + (1 + Cr^2) \sqrt{
    2 + C r^2 - C^2 r^4}\right\}}
 \end{eqnarray}
 The integration in the second term can not perform analytically due to the complexity of the expression. One can note that the mass function depends on $\chi$. The effective mass function of the system thus obtained as :
 \begin{eqnarray}
 m^{\text{eff}}(r)&=&4\pi\int_0^r \rho^{\text{eff}} r^2 dr=\frac{3}{4}\frac{Cr^3}{1+Cr^2},
 \end{eqnarray}
The effective mass of  compact star is directly proportional to its radius, as seen by the behavior of the mass function in Fig.~\ref{m11}, moreover mass function is regular at the core. In this graph, we can see that the maximum mass is achieved at the boundary of the star.
The effective compactness factor $u^{\text{eff}}$ which classifies the compact objects in different categories as normal star ($u^{\text{eff}} \sim 10^{-5}$), white dwarfs ($u^{\text{eff}}~\sim 10^{-3}$), neutron star ($10^{-1} <~u^{\text{eff}}~< 1/4$), ultra-compact star ($1/4<~u^{\text{eff}}<1/2$)
and black hole ($u^{\text{eff}} \sim 1/2$) can be expressed in terms of mass function as follows :
\begin{eqnarray}
u^{\text{eff}}=\frac{m^{\text{eff}}}{r}.
\end{eqnarray}
Furthermore, the following formula can be used to calculate surface redshift($z_s^{\text{eff}}$):
\begin{eqnarray}
z_s^{\text{eff}}&=&\frac{1}{\sqrt{1-2u^{\text{eff}}}}-1,
\end{eqnarray}
The profiles of effective compactness and surface redshift are shown in Fig.~\ref{m11}.

\begin{table*}[t]
\centering
\caption{The numerical values of central density, surface density, central pressure and relativistic adiabatic index $\Gamma$ at the center for the compact star LMC X-4 for different values of coupling constant $\chi$.}
\label{tb1}
\begin{tabular}{@{}cccccccccccccccc@{}}
\hline
$\chi$& $\rho_c$ & $\rho_s$ & $p_c$ & $\Gamma(r=0)$\\
\hline
 0& $1.08741 \times 10^{15}$& $6.89096 \times 10^{14}$ & $1.2973 \times 10^{35}$&3.02358\\
 0.25& $1.05726 \times 10^{15}$ & $6.69128 \times 10^{14}$& $1.25226 \times 10^{35}$&3.01932\\
 0.5& $1.02872 \times 10^{15}$ & $6.50285 \times 10^{14}$ & $1.211 \times 10^{35}$&3.01586\\
 0.75& $1.00167 \times 10^{15}$ & $6.32474 \times 10^{14}$ & $1.17304 \times 10^{35}$&3.01306\\
 1& $9.76004 \times 10^{14}$ & $6.15612 \times 10^{14}$ & $1.13797 \times 10^{35}$&3.01079\\
 1.25& $9.51611 \times 10^{14}$& $5.99627 \times 10^{14}$ & $1.10545 \times 10^{35}$&3.00896\\
\hline
\end{tabular}
\end{table*}

\section{Discussion}

In this paper, we present a new solution to Einstein's field equations in the $f(R,T)$ theory of gravitation, as well as a model for the compact star LMC X-4. LMC X-4 features a superorbital X-ray cycle and a 1.41-day orbital period. LMC X-4 is a two-star system consisting of a pulsar-a highly magnetised neutron star beaming X-rays - and a companion star. As these astrophysical objects are basically spherically symmetric configurations with Buchdahl metric potential, so the physical properties are more or less same for these objects. This compact star has been successfully used earlier by several researchers to develop stellar model. That is why we choose compact stars candidate LMC X-4 for our present paper. After conducting a thorough analysis of the long-term fluctuations in LMC X-4's X-ray flux, Heemskerk \& van Paradijs \cite{heemskerk1989analysis} came to the conclusion that the object contained a warped precessing accretion disc. We presented a comprehensive investigation of proposed compact star models, both analytically and graphically. The metric potentials are regular and free of any sort of singularities, as seen in Fig.~\ref{metric}. The figure shows that all of the $e^{\lambda}$ profiles coincide for different values of $\chi$, although $e^{\nu}$ assumes a higher value for larger values of $\chi$. Inside compact stellar configurations, the physical variables pressure and density are well defined. In Fig.~\ref{pp}, these parameters are graphically shown. A detailed analysis of the figures indicates that inside the stellar configurations, both pressure and density have positive definite values and are monotonically decreasing functions of the radial coordinate `r'. The vanishing nature of the pressure indicates the size of stellar configuration. It is clear from the statistics that as the value of $\chi$ increases, both pressure and density decrease. Based on our findings, the star becomes less compact as the $\chi$ value increases. Fig.~\ref{grad5} depicts the pressure and density gradients graphically.
The gradients of pressure and density vanish at the center of the star, and $\frac{d^2p}{dr}$ and $\frac{d^2\rho}{dr}$  take negative values (Fig.~\ref{rho2}) at the center, confirming that these parameters are decreasing inside the compact star. All the energy conditions are well behaved inside the stellar interior for $0\leq \chi \leq 1.5$ as shown in Fig.~\ref{ener1}. One of the most important characteristics of the stellar configuration is the causality condition. Fig.~\ref{sv} shows that the square of the sound velocity is less than $1$, demonstrating that the causality criterion is met for our model. In  $f(R,T)$ modified gravity, the relativistic adiabatic index, which is a monotonic increasing function of radial coordinate r and takes a value greater than $4/3$, guarantees the stability of the current model (Fig.~\ref{sv}). As the value of $\chi$ increases, the star becomes more stable, as illustrated in the figure, with increasing value of $\chi$. Our star models do a good job of maintaining hydrostatic equilibrium. As a result, under the impact of the active forces, our proposed structures are stable. The tables show the numerical values of several model parameters. Also, Inside compact stellar formations, the mass function and compactness factors are well behaved, as seen in Figs.~\ref{m11} for various values of $\chi$. The surface red shift is maximum at the boundary and increases monotonically as one moves from the centre to the boundary for all values of $\chi$, as shown in Fig.~\ref{m11}. One can note that the compactification factor is less than $4/9$ everywhere inside the stellar configuration and hence Buchdahl condition is well satisfied \cite{Buchdahl:1959zz}. In the absence of a cosmological constant, Buchdahl \cite{Buchdahl:1959zz} provided the upper bound for surface redshift for a spherical object as $z_s~\leq~2$, which Bohmer and Harko \cite{Boehmer:2006ye} generalized for an anisotropic spherical object in the presence of a cosmological constant as $z_s \leq 5$. Our current model meets this constraint. As a result, all of these characteristics together provide a strong justification for the stability and viability of the compact star models presented in $f(R,\,T)$ theory.

\section*{Acknowledgements} P.B. is thankful to the Inter University Centre for Astronomy and Astrophysics (IUCAA), Pune, Government of India, for providing visiting associateship. PB also acknowledges that this work is carried out under the research project Memo No: $649$(Sanc.)/STBT-$11012(26)/23/2019$-ST SEC funded by Department of Higher Education, Science \& Technology and Bio-Technology, Government  of West Bengal.\\

{\bf Data Availability Statement:} No Data associated in the manuscript

\bibliography{Buchadahl_1}

\end{document}